\documentclass[]{spie}  

\usepackage{amsmath,amsfonts,amssymb}
\usepackage{graphicx}
\usepackage[colorlinks=true, allcolors=blue]{hyperref}
\usepackage{comment}
\usepackage{academicons}
\usepackage{xcolor}
\usepackage{subfig}
\usepackage[numbers]{natbib}
\usepackage{xspace}

\usepackage{tikz,xcolor,hyperref}
\definecolor{lime}{HTML}{A6CE39}
\DeclareRobustCommand{\orcidicon}{%
    \begin{tikzpicture}
    \draw[lime, fill=lime] (0,0) 
    circle [radius=0.16] 
    node[white] {{\fontfamily{qag}\selectfont \tiny ID}};
    \draw[white, fill=white] (-0.0625,0.095) 
    circle [radius=0.007];
    \end{tikzpicture}
    \hspace{-2mm}
}
\newcommand{\orcid}[1]{\href{https://orcid.org/#1}{\orcidicon}}
\newcommand{\fdeg}{\mbox{\ensuremath{.\!\!^\circ}}}
\newcommand{\registered}{\textsuperscript{\textregistered}\xspace}

\title{An overview of stray light findings and interpretation during on-sky commissioning of LSSTCam}

\author[a]{Gabriele Rodeghiero\orcid{0000-0002-3469-9863}}
\author[b,c]{Alex Drlica-Wagner\orcid{0000-0001-8251-933X}}
\author[a,d]{Alessio Taranto\orcid{0009-0009-3271-3498}}
\author[a,d]{Luca Rosignoli\orcid{0000-0002-0327-5929}}
\author[e]{Hannah M.M. Pollek\orcid{0009-0001-3368-4539}}
\author[c]{Aashay Pai\orcid{0009-0008-9641-6065}}
\author[f]{R. Lynne Jones\orcid{0000-0001-5916-0031}}
\author[f]{Erin Leigh Howard\orcid{0000-0002-0716-947X}}
\author[g]{Sean P. MacBride\orcid{0000-0002-9514-7245}}
\author[h]{John Andrew}
\author[h]{Douglas R. Neill}
\author[e]{Travis Lange\orcid{0009-0008-0596-4489}}
\author[e]{Andrew P. Rasmussen\orcid{0009-0000-3218-9846}}
\author[e]{Aaron Roodman\orcid{0000-0001-5326-3486}}
\author[h]{Brian Johnson}
\author[i]{Elana K. Urbach\orcid{0000-0002-3205-2484}}
\author[h]{Parker Fragelius}
\author[e]{Eli S. Rykoff\orcid{0000-0001-9376-3135}}
\author[y,l]{Tomislav Vucina}
\author[i]{Christopher W. Stubbs\orcid{0000-0003-0347-1724}}
\author[m]{Robert H. Lupton\orcid{0000-0003-1666-0962}}
\author[f]{Charles F. Claver}
\author[e]{Joshua E. Meyers\orcid{0000-0002-2308-4230}}
\author[h]{Anastasia Alexov\orcid{0009-0000-7835-3963}}
\author[n]{Keith Bechtol\orcid{0000-0001-8156-0429}}
\author[m]{Lee S. Kelvin\orcid{0000-0001-9395-4759}}
\author[h]{Brian Stalder\orcid{0000-0003-0973-4900}}
\author[o]{Pierre Antilogus\orcid{0000-0002-0389-5706}}
\author[p]{Alexandre Boucaud\orcid{0000-0001-7387-2633}}
\author[q]{Aurelien Marini}
\author[e]{Alexander Broughton\orcid{0000-0001-6966-5316}}
\author[l]{Leanne P. Guy \orcid{0000-0003-0800-8755}}
\author[h]{Tiago Ribeiro\orcid{0000-0002-0138-1365}}
\author[h]{Erik Dennihy\orcid{0000-0003-2852-268X}}
\author[l]{Bruno C. Quint\orcid{0000-0002-1557-3560}}
\author[r]{Aaron E. Watkins\orcid{0000-0003-4859-3290}}
\author[l]{Alysha B. Shugart\orcid{0009-0000-6778-7168}}
\author[l,e]{Lukas Eisert\orcid{0000-0003-3918-7995}}
\author[l]{Kevin Fanning\orcid{0000-0003-2371-3356}}
\author[l]{Marina S. Pavlovich\orcid{0000-0001-5560-7051}}
\author[l,e]{Yijung Kang\orcid{0000-0002-5261-5803}}
\author[l,$\gamma$]{Hye Yun Park\orcid{0000-0002-7295-2743}}
\author[l]{Paulo Lago\orcid{0009-0005-4105-5168}}
\author[l]{Kristopher Mortensen\orcid{0000-0001-9676-5005}}
\author[l]{Paulina Venegas Salas\orcid{0009-0001-3922-9588}}
\author[l]{Minhee Hyun\orcid{0000-0003-4738-4251}}
\author[l]{Karla Peña Ramírez\orcid{0000-0002-5855-401X}}
\author[l]{David Sanmartim\orcid{0000-0002-9238-9521}}
\author[l,e]{Shuang Liang}
\author[l]{Gonzalo Aravena\orcid{0009-0006-5850-4860}}
\author[l]{Kshitija Kelkar\orcid{0000-0002-8130-3593}}
\author[l,e]{Kate Napier\orcid{0000-0003-4470-1696}}
\author[l]{Jacqueline C. Seron Navarrete\orcid{0000-0002-8303-776X}}
\author[l]{Carlos A. L. Morales Marín\orcid{0000-0003-0203-3407}}
\author[l]{Danica Žilková\orcid{0000-0002-5726-3640}}
\author[l,e]{Narayan Khadka}
\author[l]{Eric J. Christensen\orcid{0009-0001-9424-2291}}
\author[s]{Yousuke Utsumi\orcid{0000-0001-6161-8988}}
\author[$\alpha$]{Merlin Fisher-Levine\orcid{0000-0001-9440-8960}}
\author[m]{Yusra Alsayyad\orcid{0009-0008-9216-7516}}
\author[f]{Colin T. Slater\orcid{0000-0002-0558-0521}}
\author[e]{Fritz M\"uller\orcid{0000-0002-7061-4644}}
\author[l]{William O'Mullane\orcid{0000-0003-4141-6195}}
\author[t]{Enrico Giro\orcid{0000-0001-7301-8285}}
\author[u]{Rodolfo Canestrari\orcid{0000-0003-4591-7763}}
\author[v]{Guillem Homar Megias\orcid{0000-0001-6013-1131}}
\author[h]{Sandrine J. Thomas\orcid{0000-0002-9121-3436}}
\author[l,e]{Kevin A. Reil\orcid{0000-0002-2234-749X}}
\author[l]{Roberto Tighe}
\author[l]{Mario Rivera}
\author[l]{Juan Lopez}
\author[l]{Claudio H. Araya Cortes}
\author[l]{David Jiménez Mejías}
\author[l]{Hernán Herrera}
\author[l]{Freddy Muñoz Arancibia}
\author[x]{Dimitri Buffat}
\author[x]{Johan Bregeon\orcid{0000-0002-6790-5328}}
\author[z,l]{Jacques Sebag\orcid{0000-0001-9348-0290}}
\author[l]{Holger Drass\orcid{0000-0002-7790-9971}}
\author[l]{Pablo Zorzi}
\author[w]{Massimo Brescia\orcid{0000-0001-9506-5680}}

\affil[a]{\small INAF OAS, Via Gobetti 93/3, I-40129, Bologna, Italy}
\affil[b]{\small Fermi National Accelerator Laboratory, P.O.\ Box 500, Batavia, IL 60510, USA}
\affil[c]{\small Department of Astronomy \& Astrophysics, University of Chicago, Chicago, IL 60637, USA}
\affil[d]{\small Department of Physics and Astronomy, University of Bologna, Via Gobetti 93/2, I-40129, Bologna, Italy}
\affil[e]{\small SLAC National Accelerator Laboratory, Menlo Park, CA 94025, USA}
\affil[f]{\small University of Washington, Dept. of Astronomy, Box 351580, Seattle, WA 98195, USA}
\affil[g]{\small Physik-Institut, University of Zurich, Winterthurerstrasse 190, 8057 Zurich, Switzerland}
\affil[h]{\small Vera C.~Rubin Observatory Project Oﬀice, 950 N. Cherry Ave., Tucson, AZ 85719, USA}
\affil[i]{\small Department of Physics, Harvard University, 17 Oxford St., Cambridge MA 02138, USA}
\affil[l]{\small NSF-DOE Vera C.\ Rubin Observatory / NSF NOIRLab, Casilla 603, La Serena, Chile}
\affil[m]{\small Department of Astrophysical Sciences, Princeton University, Princeton, NJ 08544, USA}
\affil[n]{\small Physics Department, 2320 Chamberlin Hall, University of Wisconsin-Madison, 1150 University Ave. Madison, WI 53706-1390 USA}
\affil[o]{\small Sorbonne Université, Université Paris Cité, CNRS/IN2P3, LPNHE, 4 place Jussieu, F-75005 Paris, France}
\affil[p]{\small Université Paris Cité, CNRS/IN2P3, APC, 4 rue Elsa Morante, F-75013 Paris, France}
\affil[q]{\small CPPM, 163, avenue de Luminy - Case 902, 13288 Marseille cedex 09, France}
\affil[r]{\small Centre for Astrophysics Research, University of Hertfordshire, College Lane, Hatfield AL10 9AB, UK}
\affil[s]{\small National Astronomical Observatory of Japan, Chile Observatory, Los Abedules 3085, Vitacura, Santiago, Chile}
\affil[t]{\small INAF OATS, Via Giovan Battista Tiepolo 11, 34143, Trieste, Italy}
\affil[u]{\small INAF IASF, Via Ugo la Malfa 153, 90146, Palermo, Italy}
\affil[v]{\small California Institute of Technology, 1200 East California Boulevard Pasadena, California 91125}
\affil[z]{\small W.~M.~Keck Observatory, 65-1120 Mamalahoa Hwy, Kamuela, HI 96743, USA}
\affil[w]{\small Department of Physics ”E. Pancini”, University Federico II of Napoli, Via Cintia, 80126 Napoli, Italy}
\affil[y]{\small GMTO Corporation, 300 N. Lake Avenue, 14th Floor, Pasadena, CA 91101}
\affil[x]{\small Universit\'{e} Grenoble Alpes, CNRS/IN2P3, LPSC, 53 avenue des Martyrs, F-38026 Grenoble, France}
\affil[$\gamma$]{Department of Physics, Duke University, Durham, NC 27708, USA}
\affil[$\alpha$]{D4D CONSULTING LTD., Suite 1 Second Floor, Everdene House, Deansleigh Road, Bournemouth, UK BH7 7DU}

\authorinfo{Further author information: E-mail: gabriele.rodeghiero@inaf.it}

\begin{document} 
\maketitle

\begin{abstract}
Wide-field telescopes are intrinsically difficult to shield from unwanted stray and scattered light, while the search to identify sources of contaminating light is frequently a challenging task. The Vera C.~Rubin Observatory, which achieved its first photon with the LSST Camera (LSSTCam) on April 15, 2025, will initiate a revolutionary era for the study of dark matter, dark energy, the transient sky, the Solar System, and the Milky Way. LSSTCam will provide near seeing-limited images of the sky in six bands ($u,g,r,i,z,y$) over a $3\fdeg5$-diameter field of view, and over the course of a decade, it will execute the Legacy Survey of Space and Time (LSST). 
This work provides an overview of the dedicated stray and scattered light test campaign that has been undertaken since the start of Rubin commissioning. In particular, we highlight the processes used to characterize, model, and mitigate stray light present in LSSTCam images.
The Rubin commissioning team created a series of testing and analysis tools to track stray light artifacts from their initial discovery through reproduction with timely observations, simulation using ray tracing to identify opto-mechanical origins, and finally devising corrective actions. The complex stray light features encountered by Rubin provide a wealth of experience for the future wide-field and extremely wide-field observatories. 
This work covers the many stages of a long journey that started with conceiving an innovative and challenging optical design, followed by the engineering and system engineering efforts to build it, to finally delivering an optimized and revolutionary cutting-edge facility.

\end{abstract}


\section{INTRODUCTION}
\label{intro}  

The NSF-DOE Vera C.~Rubin Observatory \cite{Ivezic19} is the world's premier wide-area, ground-based observatory at visible and near-infrared wavelengths. Over the next 10 years, the Rubin Observatory's Legacy Survey of Space and Time (LSST) will use the LSST Camera (LSSTCam; \citep{lsstcam_ref}) to image the entire visible sky every 3 days to provide an unprecedented, deep, wide-area movie of the dynamic universe. In addition, the combined imaging from Rubin will reach unprecedented sensitivity for faint astronomical sources, delivering measurements of $\sim$20 billion stars and $\sim$10 billion galaxies over the survey duration. However, the unique, compact, 3-mirror design of the Simonyi Survey Telescope \citep{blum25} at the Rubin Observatory is especially sensitive to stray and scattered light. Stray light features occur when light follows unintended direct or indirect paths from the sky or dome to the LSSTCam focal plane. Such artifacts can include light from the Moon, bright stars, or unintended light sources on the sky.   Such artifacts contaminate all astronomical imaging to some level; however, the unprecedented low-surface brightness science that is targeted by LSST is especially sensitive to these features. In this document, we summarize work performed during commissioning of the Rubin Observatory to identify, understand, and mitigate various sources of stray and scattered light.

There are several examples of recent wide-field instruments operating both from ground and space which have had to deal with stray light effects that were identified during commissioning. The Dark Energy Camera (DECam; \cite{Flaugher_2015}), a $2\fdeg2$-diameter Field of View (FoV) imager mounted at the prime focus of the Victor M.~Blanco 4-m telescope on Cerro Tololo in Chile, underwent systematic studies of stray-light sources that highlighted the presence of arcs and variously shaped sprays of light in the proximity of bright stars located $\sim 1\fdeg 5$ off-axis \cite{Kent2013, Chang2021, Tanoglidis2022}.
The mitigation strategy involved the installation of additional plastic baffles around the edge of each filter cell to cover the gap between the edge of the filter and the cell, and the painting of the interior edges of the filter aperture changer and shutter with anti-reflective paint (Aeroglaze\registered  Z306) \cite{Flaugher_2015}. 
Similarly, OmegaCam, operating in the \textit{ugriz} filters and installed on the VLT Survey Telescope (VST), encountered stray and scattered light issues, mostly affecting flat field frames \cite{vst2020}. In this case, the issue was solved by installing additional baffles. 
The SPHEREx space mission has an off-axis three-mirror free-form optical design that delivers a wide $3\fdeg5 \times 11\fdeg5$ telecentric FoV designed to map the entire sky at near-infrared wavelengths in low-resolution spectroscopy, and stray light features have been observed from bright stars in narrow regions close to the frame edges. Outside of the instrument FoV, there is a detectable artifact from stars about 8$^\circ$ off-axis \cite{Bock_2026}. 
The instrument can flag and remove images where bright stars are near regions of susceptibility as a further mitigation. Other stray light features appear to be associated with the Moon, which is always kept away from the telescope's boresight by $>32^\circ$.
The Euclid space mission suffers from stray light contamination originating from a thruster bracket lying outside the shadow of the sun-shield and receiving direct sunlight that is scattered towards the visible instrument. The latter is protected by many layers of insulation, but given the extreme sensitivity of the instrument, the light still makes it through the insulation layer at some specific angles. At these angles, about 10\% of the observations are affected by stray light\cite{esa2023}. This feature has an impact on the optimization of the Euclid survey and constrains the orientation of each pointing in the sky. 

In this context, Rubin is particularly susceptible to stray light features, due to its combination of large étendue ($A\Omega$), compact telescope design, and all-sky survey mode of observation. Several stray light features have been discovered in correspondence with specific off-axis positions for bright stars, planets, and the Moon during the Rubin Observatory commissioning. A non-negligible fraction of these artifacts were back-traced to an incomplete dome status: the dome will have a Light Wind Screen (LWS) to restrict the acceptance angle along the elevation axis of the telescope, which is not yet operational. Many stray light artifacts are expected to resolve upon installation and commissioning of the LWS. Some of the features were expected, as originally modeled during the first stray light study for LSST, which was carried out by Photon Engineering \cite{fred_06} using FRED\registered. Other artifacts were unexpected, and they required intensive visual inspection and ray tracing simulation studies to infer and confirm their origins. This paper summarizes the on-sky stray light findings during Rubin commissioning and early operations, detailing the different test and analysis methodologies developed to study, classify, and mitigate various stray light features.

\subsection{Optical design of Rubin telescope and camera}
\label{optic_des}

\begin{figure}
    \centering
    \includegraphics[width=1\linewidth]{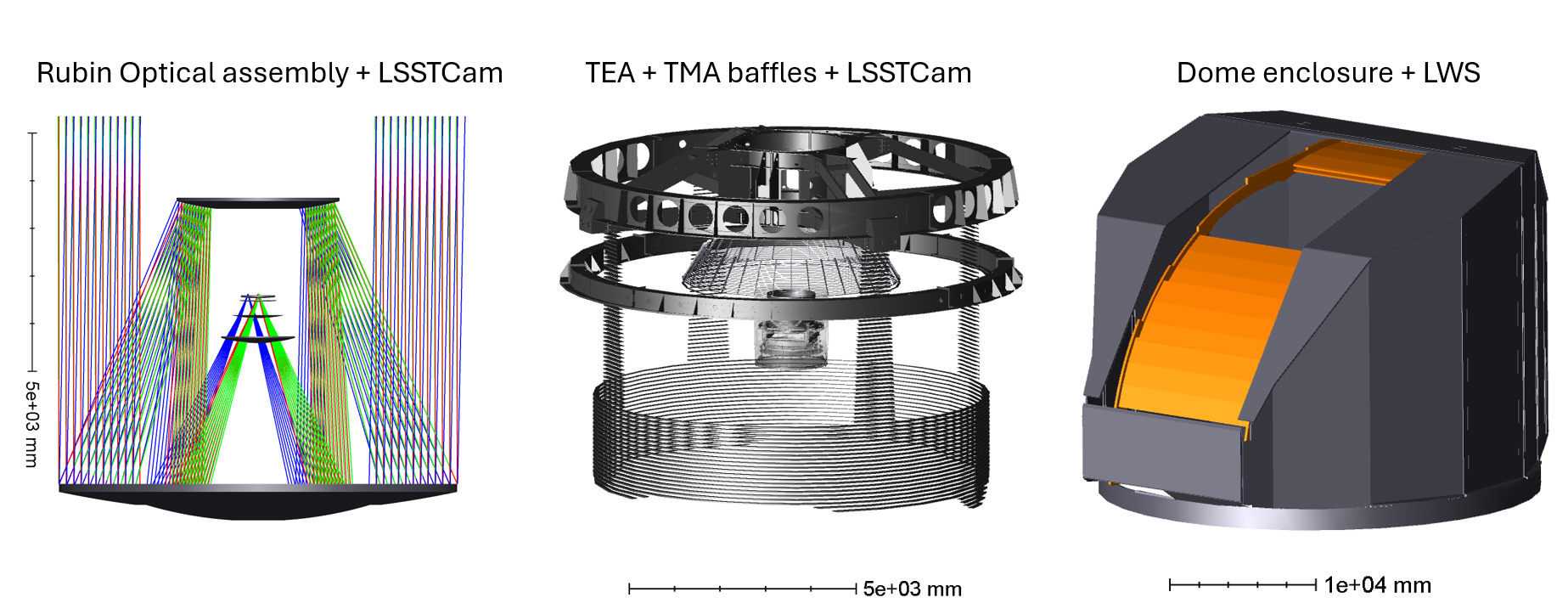}
    \caption{From left to right: optical assembly of Rubin with three powered mirrors, three lenses, and one curved filter in LSSTCam; Top End Assembly (TEA) structure and TMA baffles around M1M3 and M2; dome enclosure and Light Wind Screen (LWS, orange). The latter blocks the off-axis light coming from beyond $20^\circ$.}
    \label{baffles_original}
\end{figure}

The Simonyi Survey Telescope has a unique optical design, which is the result of the fusion of two different projects: the Dark Matter Telescope \cite{Angel2000} and the Large-aperture Synoptic Survey Telescope\cite{Tyson2001}. The project became LSST \cite{Tyson2002} (Large Synoptic Survey Telescope) in 2002 and was more recently renamed as the Vera C.\ Rubin Observatory \cite{Mountain2018}. The telescope optical configuration is a modified Paul-Baker Mersenne Schmidt, f/1.234 three-mirror system; all of the mirrors are powered, aspherical, and have non-zero conic constants. The most peculiar feature, optically speaking, is the primary and tertiary mirror cell (M1M3), with M1 and M3 cast in the same mirror blank to simplify the alignment process and ensure higher alignment stability; it allows the telescope to be very compact and therefore have a quick settling \cite{sebag16}. Like every fast optical system, Rubin is prone to stray light of multiple origins, which is not always easy to prevent or resolve. For several off-axis angles, the very compact opto-mechanical layout of the Simonyi Survey Telescope limits the possibility of inserting light baffles into the optical path without vignetting the scientific FoV.
The Telescope Mount Assembly (TMA), in addition to hosting the Rubin opto-mechanical assembly, accommodates a series of light baffles (Fig. \ref{baffles_original}) to block the primary stray light paths. An important baffle to limit the stray light from large off-axis angles ($>20^\circ$) is the LWS \cite{Marchiori:2024}, which is fully external to the TMA, and it restricts the acceptance angle of the dome shutter along the elevation axis of the telescope. At the time of writing (June 2026), the LWS installation is not yet complete, leading to an excess of stray light in the LSSTCam images.

\subsection{Initial stray light study during design phase}
\label{fred}

The first stray light study for LSST was carried out by Photon Engineering using FRED\registered in 2006 \cite{fred_06, ellis09}. The stray light analysis was performed on the full telescope and dome model, and optical surfaces were assigned with scatter models to represent the scatter due to particulate contamination and the surface micro-roughness. All non-optical surfaces were assigned a black-paint scatter model based on measurements of Aeroglaze\registered Z306. 
The stray light analysis was broken into two main tasks: point source transmittance and image 
irradiance distributions. The former is used to characterize the out-of-field stray light rejection characteristics of the telescope. The latter is used to determine the illumination across the focal plane. The study pursued the analysis of three main stray light paths: 

\begin{itemize}
    \item Direct paths to the focal plane that do not follow the imaging path 
    \item Single-scatter events from structural components that reach the focal plane
    \item Single-scatter events from all optical surfaces and 2$^{nd}$-order ghost reflections from refractive surfaces  
\end{itemize}

The stray light level was expressed in terms of Point Source Transmittance (PST), which is the ratio between detected irradiance and incident irradiance. The PST is used to determine the out-of-field stray light rejection characteristics of an optical system, but it has the limitation of not providing information about the structure of the stray light distribution on the focal plane. The simulations were run in the $r$ band filter. Only single scatter events, i.e., one scatter event from the object before illuminating the focal plane, were considered in the study, under the assumption that multiple scatter events result in negligible scattered flux.
The study highlighted a series of principal stray light sources and propagation paths:

\begin{itemize}
    \item Between 3$^\circ$ and 5$^\circ$ off-axis, scatter from the first and second lenses of LSSTCam, L1-L2, is the largest contributor to the stray light background. Within this angular range, the light enters the LSSTCam following the normal imaging path until it passes through L1. After this point, the incident light is scattered towards the detector by some of the structural components between L1 and L2.
    
    \item Between 6$^\circ$ and 12$^\circ$ off-axis, the M3 direct path is the main source of scattered light: the rays are reflected off the M3 surfaces, and they propagate into the LSSTCam, bypassing the reflection onto M1 and M2, and subsequently hit the internal structure of LSSTCam between L1 and L2.
    
    \item Between 12$^\circ$ and 30$^\circ$ off-axis, the main stray light contributions come through the M1 direct path, bypassing the reflection onto M2 and M3 and hitting the inner walls of the LSSTCam. The LWS was conceived to limit the impact of this stray light path since the TMA baffles cannot block an optical beam path coming off the inner zone of M1 at an angle of 14$^\circ$ without vignetting the scientific FoV. At the time of the study (2006), the LWS had a circular aperture (15.6m diameter), which was more restrictive than the current version, which has an angular acceptance of $\sim19\fdeg8 \times 19\fdeg8$.  
\end{itemize}

\begin{figure}
    \centering
    \includegraphics[width=0.9\linewidth]{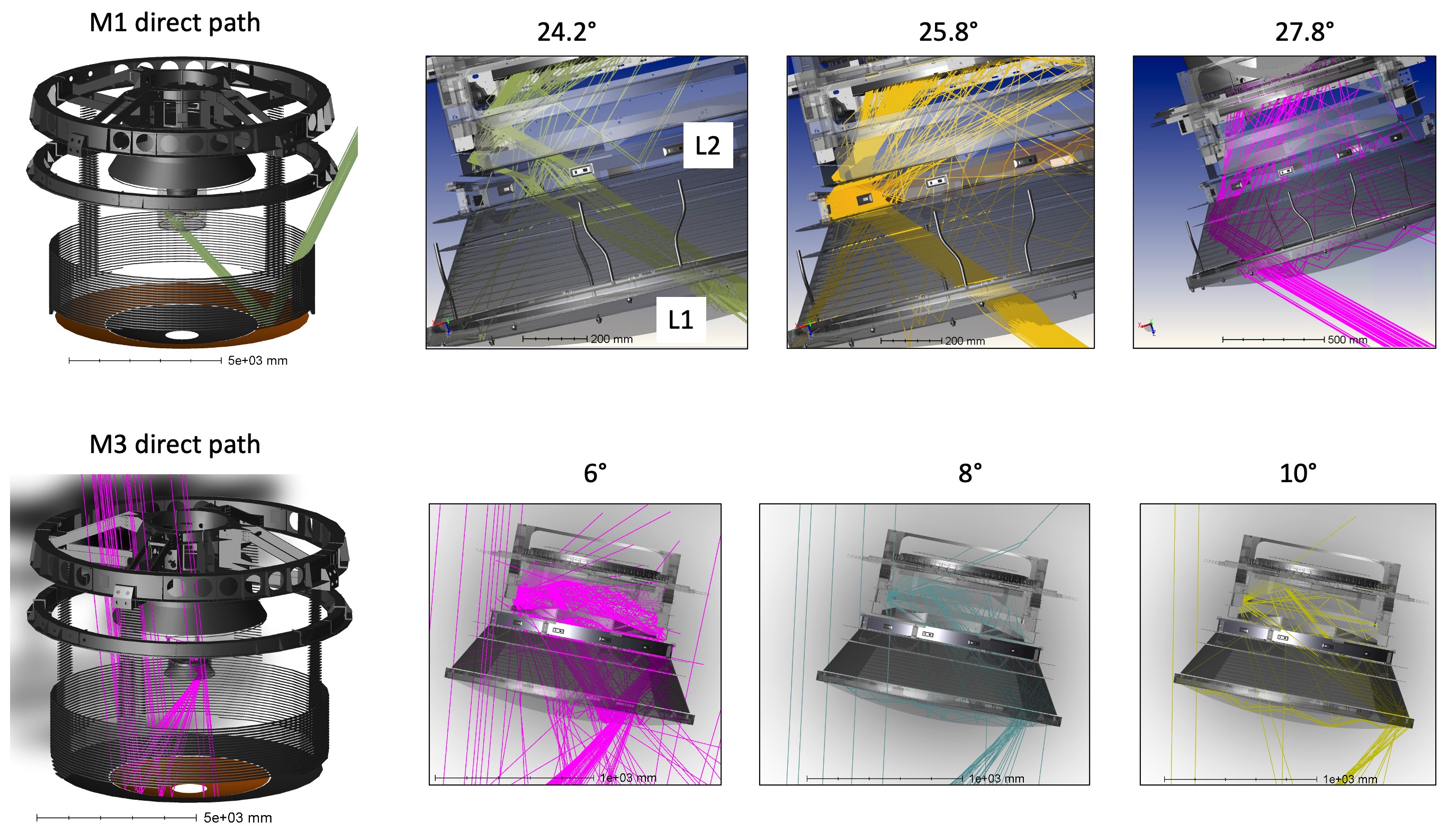}
    \caption{Non-sequential simulations in Ansys Zemax OpticStudio\registered showing the stray light bundle from the M1 (top) and M3 (bottom) direct path hitting different internal parts of the LSSTCam. As the off-axis coordinate increases, the stray light bundle is scattered off the filter autochanger area, the L2 side, and the L1-L2 baffle.}
    \label{M1_M3_direct_beam}
\end{figure}

These expectations have been confirmed by a recent stray light analysis conducted in a non-sequential scenario in Ansys Zemax OpticStudio\registered as shown in Fig. \ref{M1_M3_direct_beam}. 

The PST calculations showed that the proposed baffle configuration using the LWS and two baffle rings on the telescope is sufficient to block the direct illumination of the focal plane from sources outside of the field of view for on-sky sources. The critical and illuminated object calculations derived from the backward light propagation showed that there are regions of the focal plane that can see beyond the 
M1M3 cell, into the dome itself, which could be an operational issue. A real image of the illuminated objects is provided in Fig.~\ref{M2_baf_M1_st} of Section \ref{dome_mes}.
The M1M3 cell and the Top End Assembly (TEA) ring piers were identified as potential stray light sources, with the recommendation of equipping them with light baffles as effectively implemented by the Project (Fig.~\ref{baffles_original}).
The proposed telescope baffle configuration was proven to be less effective at blocking single-level scatter paths from the telescope and camera themselves. For angles in the outskirts of the FoV, scattering from structural components between L1 and L2 is the most critical, and since the telescope has no intermediate field stop (direct imaging system), very little can be done to block these paths. The L2 edges, bevels, and support pads were identified as a potential stray light source. Following this study, the Rubin project issued a total of 47 requirements related to stray light prevention and mitigation (see Section \ref{se_stray}).

\subsection{Serendipitous stray light findings}
\label{sere}

\begin{figure}[!t]
    \centering
    \includegraphics[width=1\linewidth]{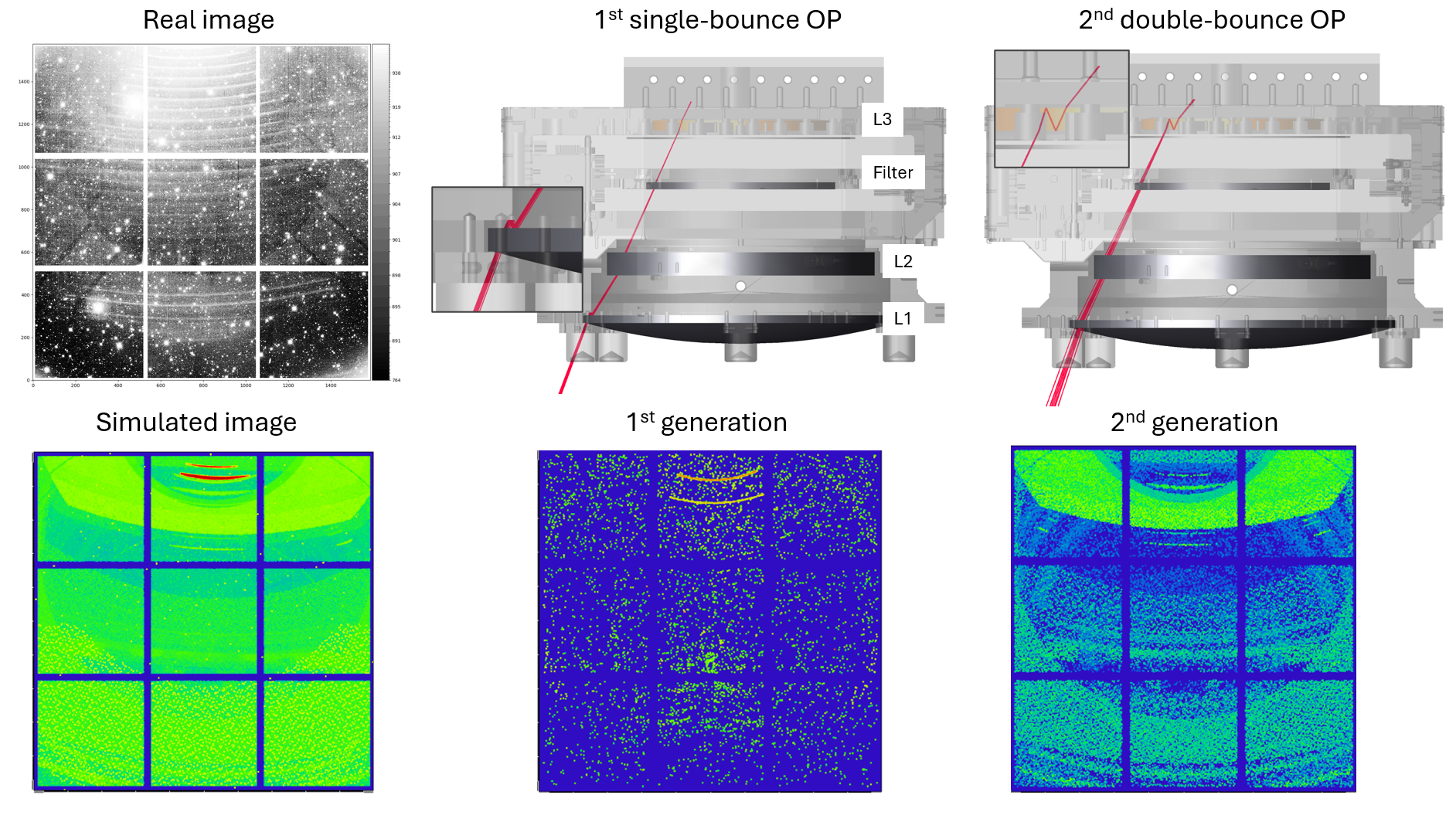}
    \caption{Top-left: Focal plane mosaic image acquired with ComCam during a dedicated stray light test using $\eta$ Columbae placed $0\fdeg47$ off-axis: a series of spurious reflections (arc-shaped) is spread over the focal plane and superimposed on a large sequence of ghosts from the star. Top-centre: Example of single-bounce (1$^{st}$ generation stray light) optical path (OP) combined with a non-optical reflection. Top-right: double-bounce (2$^{nd}$ generation) OP. Bottom, from left to right: simulated images of the overall light distribution, and of the 1$^{st}$ and 2$^{nd}$ generations of stray light patterns.}
    \label{comcam_stray}
\end{figure}

Starting from engineering first light with the Rubin Commissioning Camera (ComCam) \cite{2026arXiv260323786V}, some level of stray light was detected associated with bright astronomical objects (e.g., stars and planets). ComCam, with its field of view of $\sim$0.5 deg$^2$ (40'×40'), covered only $\sim$5\% of the LSSTCam focal plane, but it was an extremely useful tool to initiate stray light investigations on sky. Both the serendipitous stray light findings during engineering testing and scientific observations, and targeted tests carried out with ComCam, helped develop a team approach toward addressing stray light issues. A few major differences between ComCam and LSSTCam are worth mentioning: i) the significantly smaller FoV of ComCam prevented many stray light edge effects that were subsequently observed with LSSTCam; ii) a light baffle mounted in front of ComCam provided stray light rejection for off-axis ray angles $>$1 degree \cite{Stalder20}, iii) ComCam had flat filters while LSSTCam has curved substrates, iv) the ComCam primary lens (L1) was uncoated, the L2, L3 were coated with an anti-reflection (AR) MgF$_2$ coating, while LSSTCam has a more sophisticated broadband AR coating. Despite these differences, the experience with ComCam allowed the development of an initial set of testing methodologies that were invaluable preparation for the commissioning campaigns with LSSTCam, and they led to the formation of a dedicated research group within Rubin with a wide spectrum of expertise. During ComCam on-sky testing, one main stray light artifact was observed \cite{sitcomtn149} in the form of arcs crossing the instrument focal plane that were visible when the telescope was pointing near bright objects, as shown in Fig.~\ref{comcam_stray}. The features come from multiple ghost reflections and ghost generations of the ComCam optics. Fig.~\ref{comcam_stray} shows an example of the stray light from the 1$^{st}$ and 2$^{nd}$ generation of ghosts and non-optical reflections from the opto-mechanical assembly of the ComCam.  

The ray tracing simulations can reproduce some of the stray light arcs observed in the on-sky frame even though a full match is missing: simulations retrieve fewer arcs than the observed pattern. The arcs originate both from 1$^{st}$ and 2$^{nd}$ generations of ghosted stray light photons; in the single-bounce ghosts, a non-optical reflection is also involved in the OP. The arcs visible at the top of the image come from a reflection onto the ComCam inner sides and a ghost reflection onto the L1 (Top-centre Fig. \ref{comcam_stray}). The bottom arcs in the central and bottom of the image come from a double-bounce ghost reflection occurring inside the ComCam optics (Top-right Fig. \ref{comcam_stray}). The number and sharpness of the arcs depend on the angular offset of the bright star with respect to the telescope boresight angle. At the time of observation, the telescope pointing model and its offset were not yet at the optimized precision, leading to small positional deviations of the bright source with respect to the commanded position.

\section{Observational techniques for stray light investigation}
\label{emp_obs}

During the Rubin on-sky commissioning period (Oct.~2024 -- Sep.~2025)\footnote{ComCam on-sky Oct.~2024 -- Dec.~2024, LSSTCam on-sky commissioning Apr.~2025 -- Sep.~2025.}, the stray light test campaign has been organized in two subsets: tests conducted on-sky and in-dome testing using the calibration light sources. Both testing options had to negotiate access to telescope time through the general prioritization of the different subsystems testing: AOS, image quality, science verification, LSSTCam, calibration, thermal system, etc. Access to on-sky time was generally granted during nights of poor seeing conditions, while in-dome testing was generally performed on nights with bad weather. At the time of writing, $\sim$38 test cases have been issued and executed for conducting stray light studies with Rubin, cumulatively accounting for more than 14 hours on-sky and 19 hours in-dome. 

Additionally, an extensive visual inspection campaign has been carried out by the stray light team since the first astronomical photon received by LSSTCam in Apr.~2025. This consisted of regular visual inspections of focal plane mosaic images that had been corrected for instrument signatures but not background-subtracted. On-sky and calibration images were inspected on a daily to weekly cadence. Stray light artifacts were identified in the images and logged. Subsequently, the image metadata was augmented with information about the location of the dome, locations of the Moon and bright stars, and other potentially informative information. These data served an important purpose in understanding occurrence rates and plausible light sources. As of the writing of this document, $>$116,000 LSSTCam images have been visually inspected in this way.

\subsection{Empirical measurements on-sky}
\label{onsky_mes}

\begin{figure}[!h]
    \centering
    \includegraphics[width=1\linewidth]{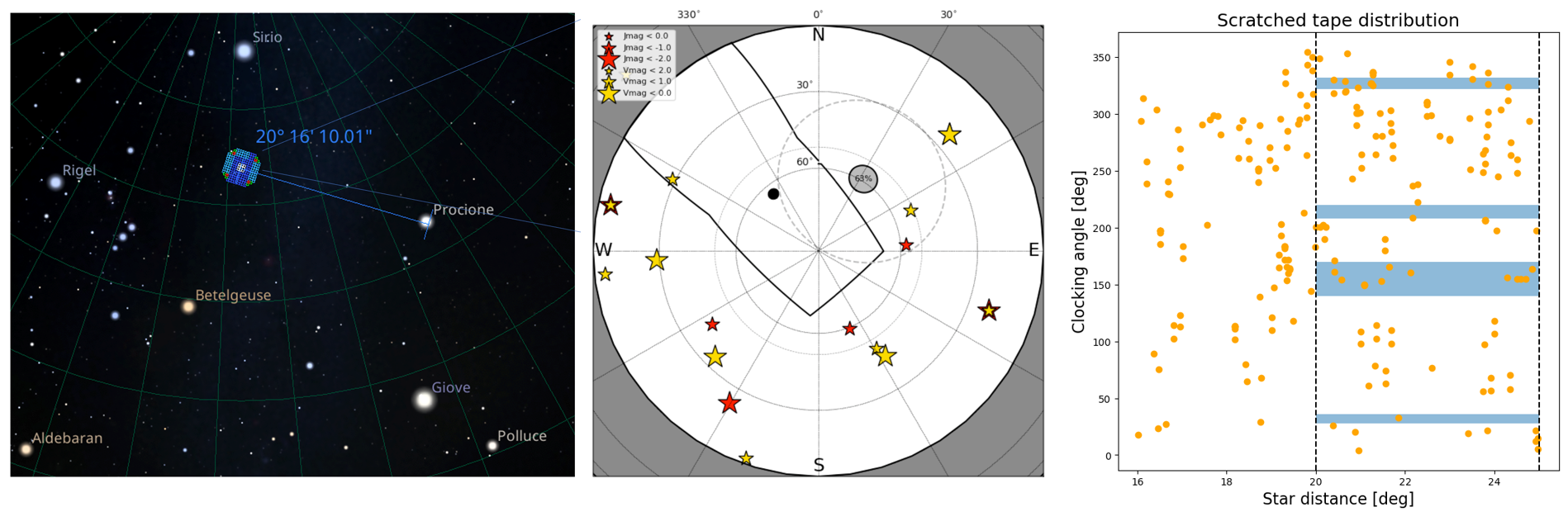}
    \caption{Examples of diagnostic figures for identifying stray light at Rubin Observatory. Left: Rubin-LSSTCam pointing projection for a specific visit on sky. A custom Python script developed by the Rubin stray light team calculates the angular distances of the bright objects and produces a \texttt{file.ZPL} to import the relevant fields and LSSTCam rotation coordinate in Zemax. Centre: Metadata about the relative locations of bright stars, the moon, and the orientation of the dome slit are augmented to the exposure list. Right: example of a diagnostic tool that correlates the LSSTcam rotator clocking angle with the celestial object off-axis coordinates and angular strips (blue) of stray light visibility for the scratched-tape feature (see Section \ref{tape}).}
    \label{astrometrization}
\end{figure}

The stray light test cases carried out on-sky were performed following different observational strategies that were strongly dependent on the specific stray light feature under study. All the tests share the need for a bright source to trigger the production of the stray light feature; in most cases, we used bright stars (mag $<$ 2) within the Rubin photometric bands, more rarely Saturn and the Moon. The test preparation and execution obey the Rubin Observing Blocks and Programs used during the on-sky commissioning. The system relies on the Jira\registered software tool to track tasks, bugs, and developing projects, historically utilized by the Project for test preparation and execution, and the requirement verification \cite{Selvy2018}. Most of the test cases were executed by running an automated procedure embedded in a JSON file deployed in the LSST Observing Visualization Environment (LOVE), which is the graphical software interface used by the Observatory to monitor the status of the telescope and its control systems \cite{Aranda2026}.
Automated scripting has the clear advantage of being executed in a reliable and repeatable way as opposed to manual executions. The discriminant between manual and automated procedures was often the length of the test: for short and simple sequences, the tests were conducted by the telescope operators and observers team, while more complex and longer sequences were fully automated, with scripts previously tested on the Rubin Base Test-stand, a virtual copy of the LOVE system providing also virtual telemetry data. The majority of the stray light tests involved a sequence of telescope pointing and camera lateral and/or rotational offsets to place and raster a bright object in specific off-axis positions where stray light features were observed or suspected to appear. A large fraction of the observed stray light features were discovered serendipitously while performing other engineering tests or scientific observations. At its first observation, the feature was named and included in a catalog of stray light features that the Rubin stray light team maintains daily. Repeated occurrences were also recorded for statistical analysis. The main stray light discoveries occurred in a parasitic fashion with the Feature Based Scheduler (FBS) testing for the development of the approach to achieve the all-sky multi-year LSST and the Active Optics System (AOS) development and commissioning. Upon every new stray light feature, and for some repeated observations, the astrometry of the bright objects within and around the LSSTCam FoV was derived using a Python script that takes the pointing information from a Rubin-Stellarium plug-in \cite{stellarium} and estimates the radial distance and position angle of the brighter objects within an arbitrary distance from the telescope boresight (Fig.~\ref{astrometrization}). The script retrieves a series of angular off-axis coordinates that are imported into the Zemax Rubin optical model through a macro (\texttt{file.ZPL}). The sequential model is turned into a non-sequential one, where scattering properties of optical and non-optical objects can be modeled as described in Section \ref{optomech} and a series of ray tracing simulations is performed to assess whether the stray light artifact can be reproduced.

\begin{figure}
    \centering
    \includegraphics[width=1\linewidth]{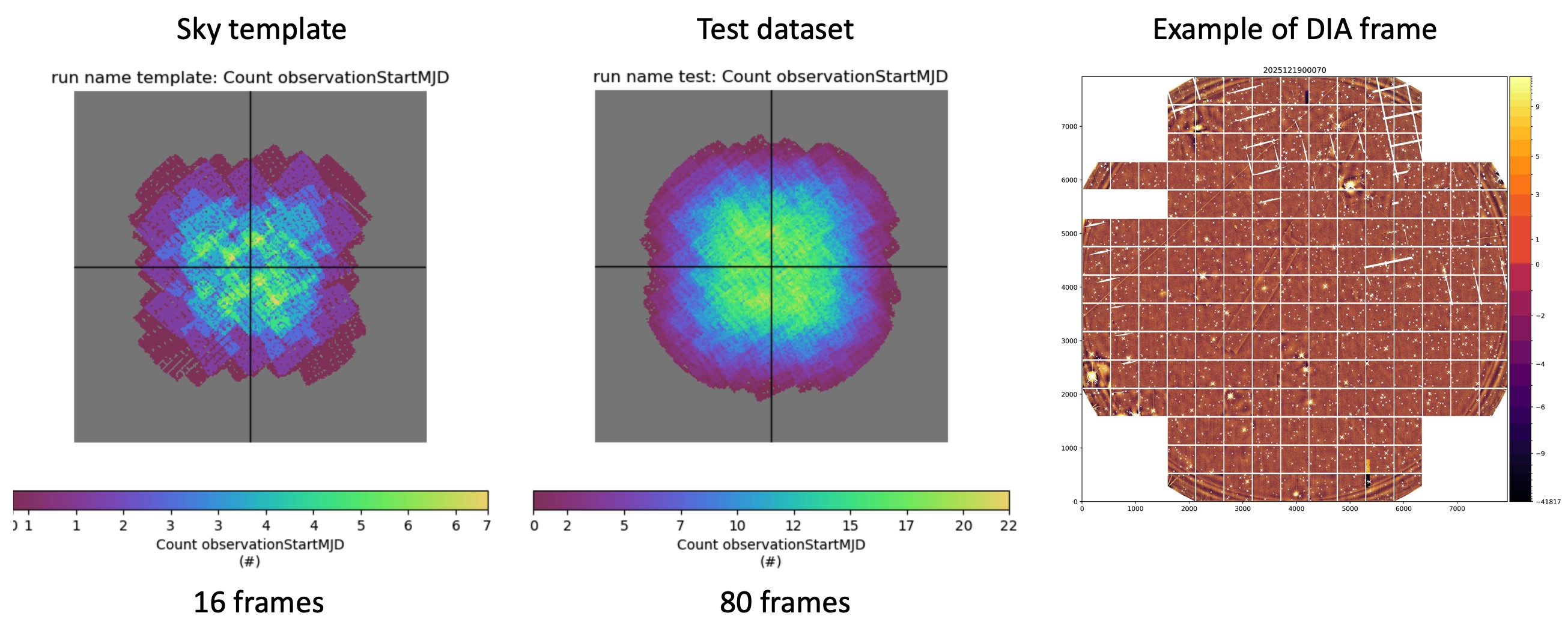}
    \caption{Difference Imaging Analysis (DIA) has been utilized for some stray light searches where the sky area to patrol was large. Two sets of data are needed: one to build an image template (left) and the other to retrieve a densely dithered image dataset (centre) from which the template is subtracted for finding the variable objects and the stray light features. Right: example of DIA frame.}
    \label{dia}
\end{figure}

In many cases, the study required designing a dedicated test case on-sky to acquire additional image frames and assessing different hypotheses for the potential offenders. To spot faint features, we also used the LSST Difference Imaging Analysis (DIA) \cite{Liu_2024} algorithm, which was developed to detect the photometric variability in the LSST. 
This technique requires two sets of data, one to build an image template and the other to retrieve a densely dithered image dataset from which the template is subtracted to find the variable objects. Using a movable bright source such as a planet, we could build a template of a patch of the sky without the stray light source, and at a second epoch, acquire the dithered data when the source was at a relevant off-axis range from the LSSTCam FoV. Since the stray light is triggered by the bright source that is only contained in the test dataset, the DIA removes all the static objects, leaving the moving, variable, and the stray light features. An example of a DIA frame is reported in Fig. \ref{dia}.

\subsection{Empirical measurements in-dome}
\label{dome_mes}

\begin{figure}
    \centering
    \includegraphics[width=0.6\linewidth]{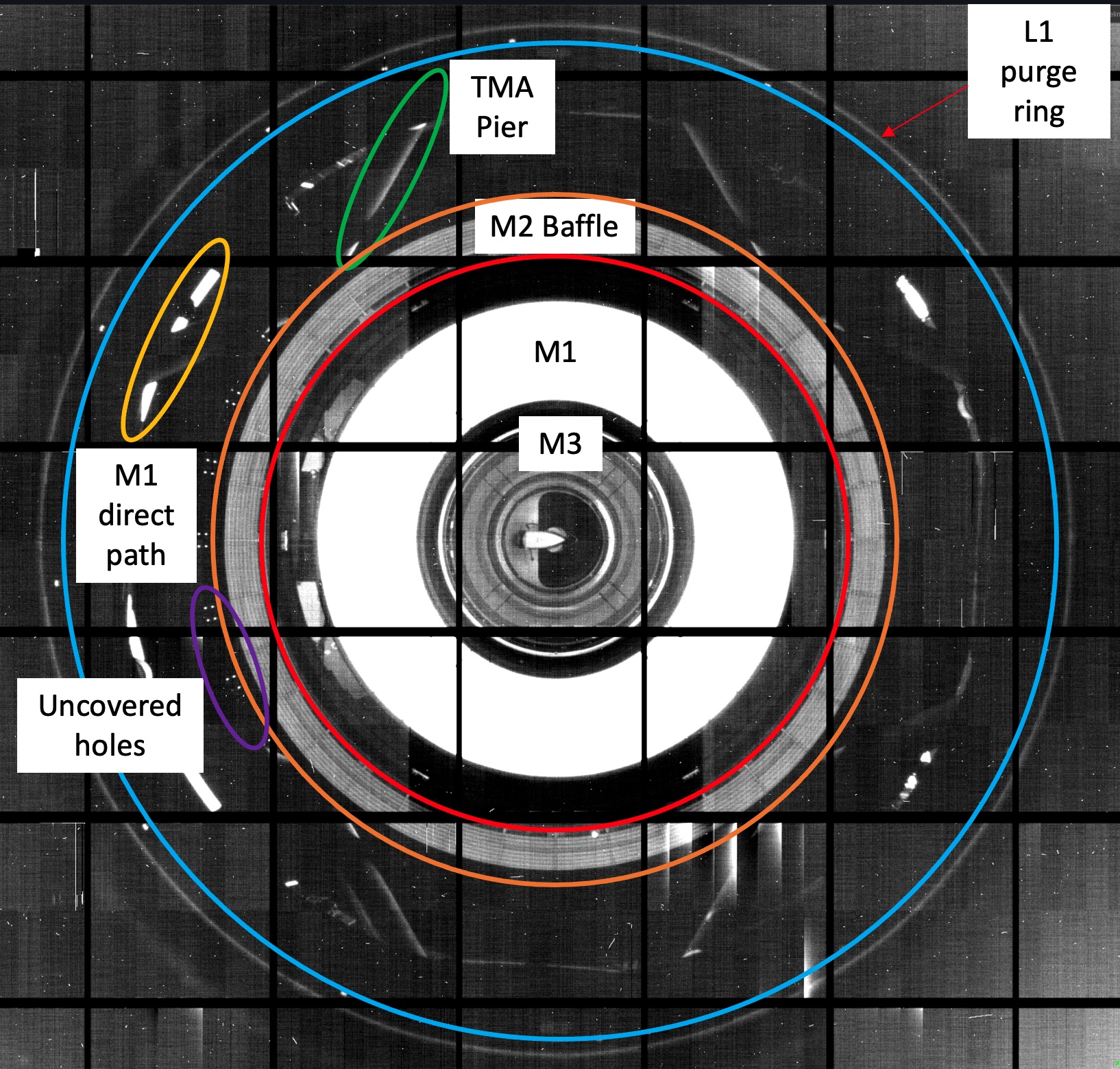}
    \vspace{1em}
    \caption{Image obtained during dawn twilight with the stenopeic (pinhole) technique for imaging the Rubin entrance pupil and the TMA opto-mechanics. A few stray and scattered light features are visible: i) the M2 baffle (red annulus) shows a higher reflectance than other TMA mechanical components due to its black anodization; ii) some light leaks (yellow ellipse) from the M1 direct path (Section~\ref{tape}); iii) some uncovered holes in the TEA baffle (purple ellipse), iv) ambient reflections off the TMA piers (green ellipse); v) reflections off the L1 purge ring (blue circle). The laser tracker mounted in the central hole of M1M3 can be seen in the centre of the scene.}
    \label{M2_baf_M1_st}
\end{figure}

In-dome studies of stray light were carried out regularly during bad weather nights. The Rubin dome is not light-tight, and it was constantly busy in the day during telescope commissioning, preventing any daytime testing for stray light. 
An interesting diagnostic tool utilized for stray light studies is stenopeic imaging, or pinhole imaging, which uses a pinhole as an imaging aperture instead of physical optics. The pinhole produces a wide-field fixed focus scene onto the camera detector plane as shown in Fig.~\ref{M2_baf_M1_st}. These image frames are acquired over the full wavelength range since the pinhole mask is deployed in substitution of a filter, either using the dome lights or a sky flat. This type of image has proven to be very useful for understanding the optical propagation paths of various stray light features. This technique was critical to understanding the origin of the \textit{scratched tape} stray light feature described in Section \ref{tape} \cite{wagner2026}.     

The Collimated Beam Projector (CBP) is a dome calibration tool to measure the telescope+LSSTCam transmission with collimated light at different wavelengths in the range of the LSST. This tool can better mimic a stellar wavefront as compared to flat-field-based instruments, allowing for more precise handling of ghosting and filter angle-of-incidence dependence \cite{Mondrik_2023}. Since the CBP can deliver a collimated, tuneable-wavelength beam, it was used intensively for in-dome stray light testing. The collimated beam projected at the telescope pupil, when utilizing the CBP mask with a single, 1\,mm pinhole, has a diameter of $\sim$30\,cm; this significant geometrical difference with respect to the light footprint of a stellar object requires precise co-alignment and off-axis tilting to produce stray light from the component under study. The CBP was used to model the level of scattered light from the M2 baffle at different off-axis angles at different wavelengths as described in Section~\ref{baffle_m2} and to refine the angular visibility window of the \textit{scratched tape} stray light artifact as described in Section~\ref{tape} \cite{wagner2026}. The CBP was also used to model the LSSTCam optical ghosts as reported in \cite{Pai2026}.

\section{Ray tracing simulations, interpretation \& mitigation actions}
\label{ray_trace}

\subsection{Opto-mechanical modeling}
\label{optomech}

\begin{figure}[!h]
    \centering
    \includegraphics[width=0.6\linewidth]{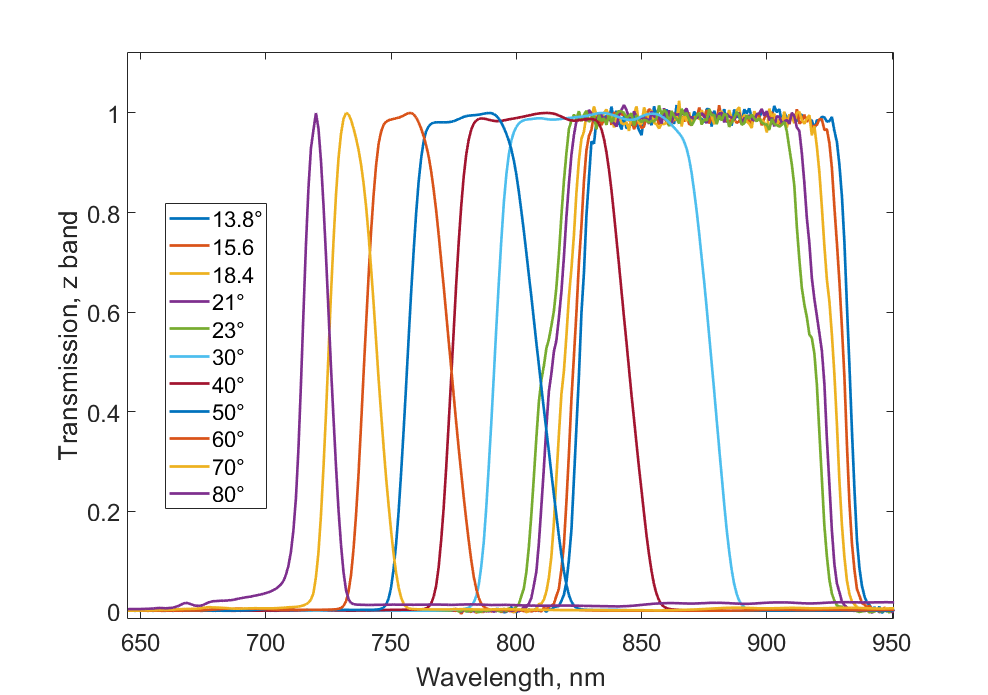}
    \caption{Example of filter transmission profile ($z$ band) measured with the Cary 5000 spectrophotometer from Agilent Technologies. Rubin delivers an f/1.2 beam to LSSTCam, so the dielectric coating of the filter is designed to operate in the range $0^\circ$--$23^\circ$, while its performance degrades quickly at higher angles of incidence by blue-shifting and narrowing the bandwidth. The data within the nominal angle of incidence range were provided by the vendor.}
    \label{z_transmission}
\end{figure}

Ansys Zemax OpticStudio\registered, used in its non-sequential mode, has the capability to import and handle mechanical format files such as STEP, IGES, STL, or SAT, which allow direct representation of telescope, camera, and dome non-optical components. This feature is not available in Batoid, which is a C++/Python optical ray tracer that models the geometric optics to characterize the optical performance of survey telescopes such as Rubin, Subaru-HSC, Blanco-DECam, and Mayall-DESI
\cite{batoid19}. Batoid played a key role in the training of the AOS algorithms and in the investigation of the \textit{scratched tape} light path (Section~\ref{tape}; \cite{wagner2026}), but it could not be used for systematic stray light investigation purposes that required detailed physical modeling of the non-optical components of the telescope, camera, and dome. Most of the stray light modeling for Rubin was thus carried out using an Ansys Zemax OpticStudio\registered non-sequential scenario.  The filter string capability of the software applied to the single-ray optical path analysis proved to be of great aid in highlighting the origin of certain stray light features. Strings allow for isolating and visualizing the rays that undergo specific propagation and scattering trajectories \cite{string_zem}; this analysis capability allows the user to select specific rays that hit or skip specific objects, undergo scattering phenomena, or are ghosted.
This modeling approach has the limitation of being computationally time-consuming and lengthy. The heaviness of the simulations is mainly due to the large size of the CAD files, the large quantity of rays shot for the analysis, and the scattering models attached to the optical and non-optical surfaces. The selection and import of specific CAD items were rationalized in relation to the likelihood of being exposed to either direct or scattered light, which can result in the potential loss of some stray light feature detection/reproduction. Also, the choice of the scattering and reflectance/transmittance profiles for the different optical and non-optical components suffers from some obvious, intrinsic limitations. Some telescope optical and non-optical surface reflection spectra were measured with a portable reflectometer, CT7 by OPO, over a range of wavelengths (365\,nm, 404\,nm, 464\,nm, 522\,nm, 624\,nm, 760\,nm, 970\,nm). The filter transmission curves in the nominal operating Angle of Incidence (AoI) range were available from the manufacturer, and they were measured over a wide AoI range as shown in Fig.~\ref{z_transmission}. 

 The TMA baffles, which have their inner surfaces painted with Aeroglaze\registered Z306, are represented with a gray (no wavelength dependence) profile with a residual reflectance of $\sim$4\%. Two main families of simulations were used for studying most of the stray light features: 

\begin{itemize}
    \item trace several rays with the relevant Rubin opto-mechanical components; look at the detector image and isolate the feature (if visible) at the detector plane corresponding to the observed artifact using a ray filter string that crops the region of interest; visualize this subset of rays in the 3D layout of the telescope to assess the stray light optical path and origin.
    \item trace several rays with the relevant Rubin opto-mechanical components; visualize the light scattered from a specific, single object of the Rubin opto-mechanical model onto the detector plane; if the feature resembles the observed artifact, visualize the subset of rays in the 3D layout of the telescope to assess the stray light optical path and origin.
\end{itemize}

\begin{figure}[!h]
    \centering
    \includegraphics[width=1\linewidth]{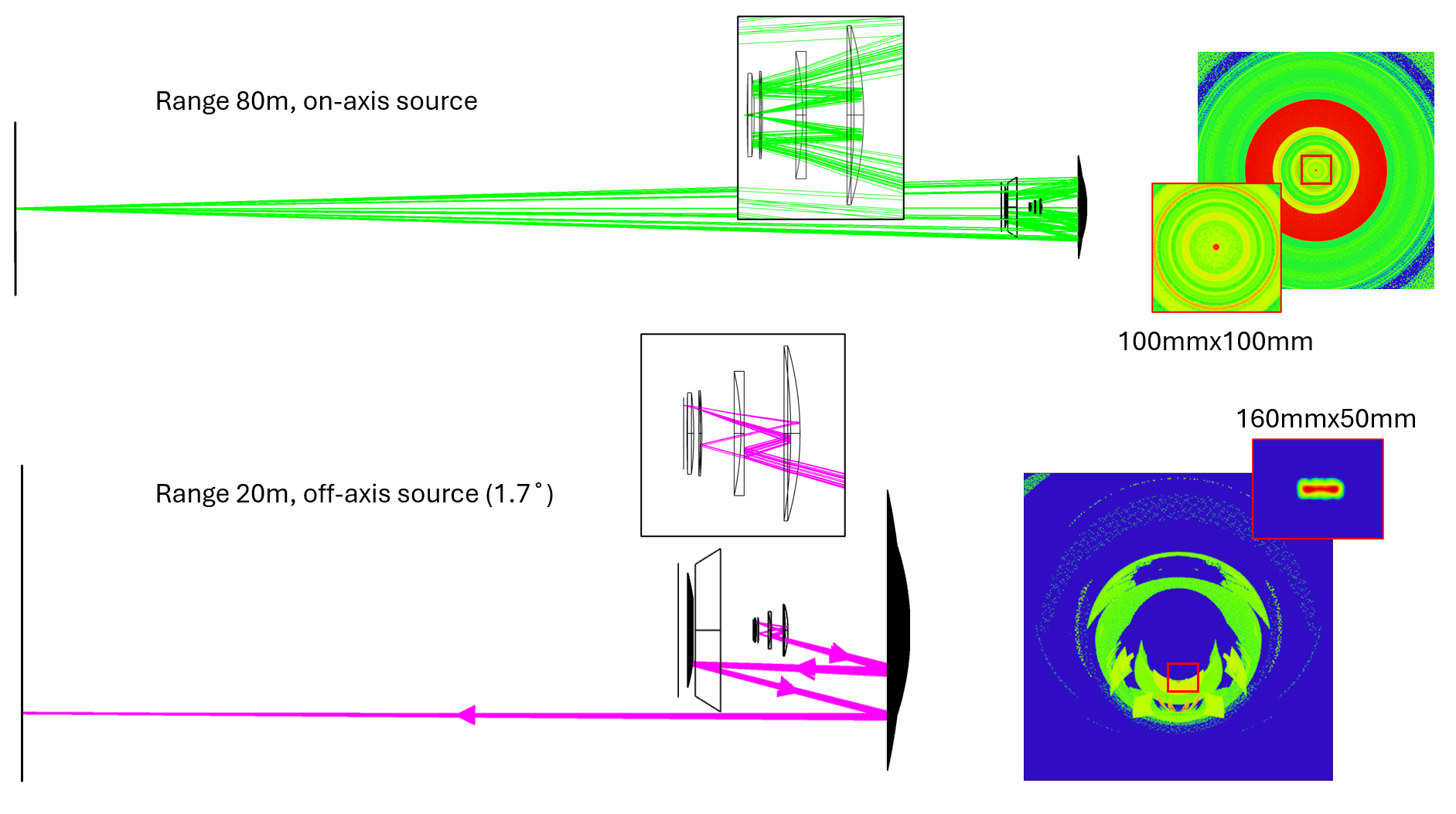}
    \caption{Top: Example of zones of convergence of the light beam found with the backward propagation path, from detector to sky. The on-axis source is ghosted by the L1 and then the L3, and it is focused $\sim$80m upstream of the M1M3 cell. Such a stray light focus is fully outside the dome, and it is therefore not critical. Bottom: A light source 1.7$^\circ$ off-axis gets ghosted by the L1 and then the L2, and gets partially focused $\sim$20m upstream of the M1M3 cell; this distance corresponds to the dome shutter position, and it could represent an issue for the creation of partially structured stray light at the detector plane.
    }
    \label{back_path}
\end{figure}

\subsection{Forward and backward light propagation, predictive modeling and reproducibility}
\label{forw_back}

The common approach to performing ray tracing simulations resembles the real world: the light propagates from the sky to the focal plane. As reported in the previous paragraph, attempts to replicate the observed stray light features in simulations have followed this \textit{forward} modeling. An alternative and complementary approach is to flip the direction of light propagation into a \textit{backward} propagation. In this case, the light source is placed at the system detector plane, and the rays are propagated backward towards the sky to search for potential zones of convergence of the optical beams within the telescope and dome volume. These zones of convergence are produced by the optical ghosts within the refractive elements, and they are governed by the conjugated points law. If such zones exist, they can lead to structured stray light features at the telescope focal plane. The light sources are propagated at different points in the FoV to cover the whole linear size of the detector plane, as shown in Fig. \ref{back_path}. 

For each field point, we scan the volume enclosed by the telescope (range $\sim$7\,m) and the dome enclosure (range $\sim$20\,m). The volume tomography is done by moving a large detector plane (14\,m $\times$ 14\,m) in steps of 100\,mm, within the TMA and of 1\,m within the dome. Moving the source off-axis explores different focal points of the optical ghosts because all the telescope mirrors and the L2 of the LSSTCam have aspherical components, which focus the ghosts at different points. Examples of zones of convergence of the light beam found with this technique are shown in Fig.~\ref{back_path}. Some zones of convergence of the light beam are found with the backward propagation path: an on-axis source is ghosted by the L1 and then the L3, and focused $\sim$80\,m upstream of the M1M3 cell. This focal plane is fully outside the dome, and it therefore does not constitute a criticality. Another zone of convergence (approximate size 160\,mm $\times$ 50\,mm) is materialized by a light source 1.7$^\circ$ off-axis which is ghosted by the L1 and then the L2 and gets partially focused $\sim$20\,m upstream of the M1M3 cell; this distance corresponds to the dome shutter position, and if found to be visible in the real images, it could represent an issue for partially structured stray light at the detector plane. Dedicated tests to confirm the visibility of such stray light paths have not yet been performed on-sky, but the technique and more detailed simulation results are collected in another paper (Rodeghiero et al., in prep.).

\subsection{Examples of stray light findings \& baffling solutions}
\label{arti_baffle}

In the following paragraphs, we report some examples of stray light features observed in the Rubin image frames; this list is not complete, but this subset allows us to highlight both the testing and modeling techniques put in place for this investigation task and the conceived mitigation strategies. A more complete census of the observed stray light artifacts is reported in Table \ref{tab:rates}.

\subsubsection{The Scratched Tape Feature}
\label{tape}

\begin{figure}[!h]
    \centering
    \includegraphics[width=1\linewidth]{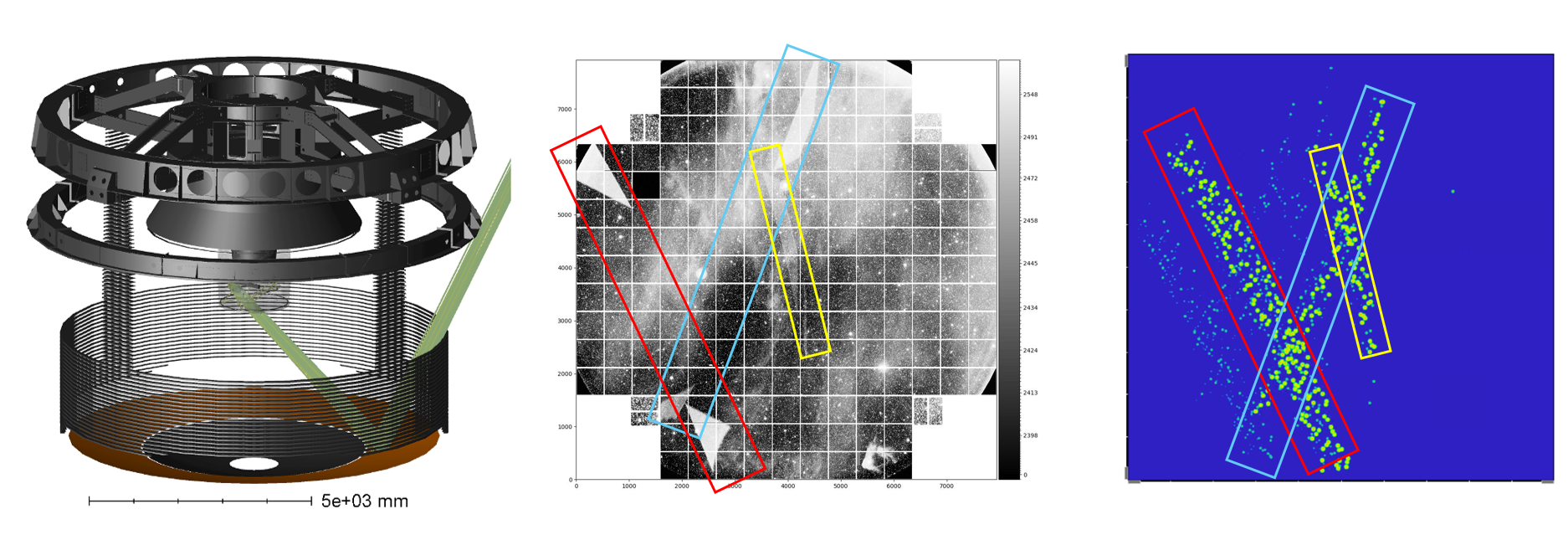}
    \caption{Left: The stray light optical path that generates the {\it scratched tape} feature is an M1 direct path, which passes between the mid-ring baffle and the M1M3 baffle. Centre: On-sky example of the different morphologies of the scratched tape feature originating from 3 bright stars ($\alpha$\,Centauri, Hadar, and Gacrux) located $\sim$20$^\circ$ off-axis. Right: Feature reproduction with the Ansys Zemax OpticStudio\registered non-sequential model.}
    \label{scratch_feature}
\end{figure}

The most frequent and impactful stray light feature observed in the LSSTCam image frames during on-sky commissioning is the so-called \textit{scratched tape} feature \cite{wagner2026}. This feature originates from an M1 direct path that skips reflections on M2 and M3 as shown in Fig.~\ref{scratch_feature}. The LWS would naturally shield this stray light path from the sky, but due to the delay in the dome completion, this feature was visually apparent in approximately 7.4\% of the image frames, with an off-axis range of visibility between $\sim$19\fdeg8 and 22$^\circ$. As can be seen in Fig.~\ref{scratch_feature}, multiple \textit{scratched tape} features can overlap in a single image when an asterism of bright stars (mag $<$ 2) resides in the relevant off-axis range.

The origin of the feature was discovered thanks to the pinhole imaging technique described in Section \ref{dome_mes}, and a $\sim$220\,mm annular extension of the mid-level baffle ring was developed to block this stray light. The baffle extension was installed between February and April 2026, and the LSSTCam images no longer suffer from this prominent stray light feature. A more detailed description of the \textit{scratched tape} phenomenology and its mitigation strategy is reported in Drlica-Wagner et al. (2026) \cite{wagner2026}.

\subsubsection{L3 Angel Wings \& Horseshoe}
\label{angel}

\begin{figure}[!h]
    \centering
    \includegraphics[width=0.8\linewidth]{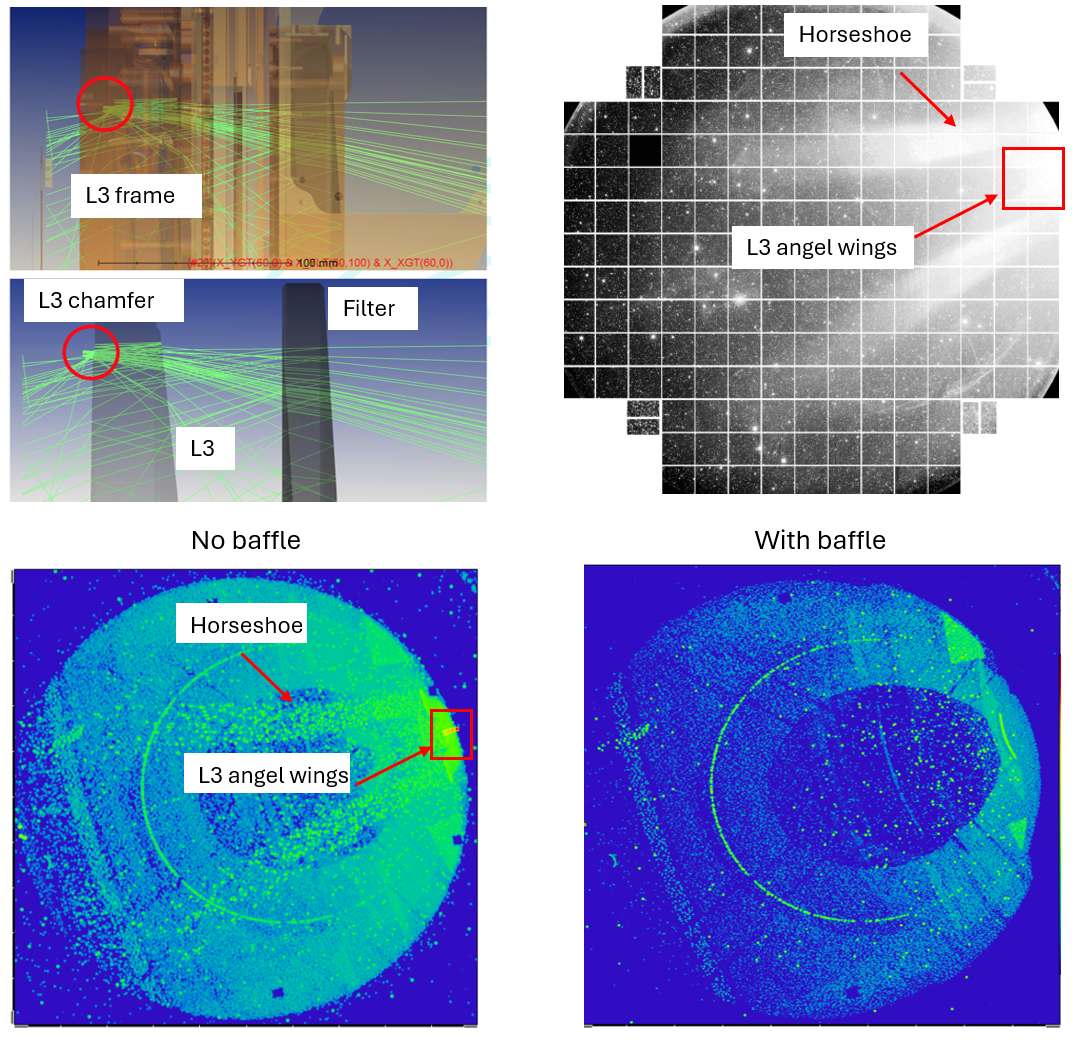}
    \caption{Top-Left: The stray light optical path that generates the {\it Horseshoe} and the {\it L3 Angel Wings} artifacts is linked to a refraction by the L3 chamfer and a reflection onto a metal edge in the cold side of the L3. Top-right: {\it Horseshoe} (most prominent) and {\it L3 Angel Wings} (smaller) stray light features materialized by Canopus placed at $2\fdeg02$ off-axis. Bottom-left: The ray tracing simulations with Ansys Zemax OpticStudio\registered reproduce both the artifacts faithfully. Bottom-right: Both features could be damped by an additional baffle to be mounted in front of L3.}
    \label{horse_feature}
\end{figure}

The \textit{L3 Angel Wings} \& \textit{Horseshoe} are stray light artifacts that originate in the immediate outskirts of the LSSTCam FoV. They appear and reach their maximum spatial extension for off-axis ranges between $1\fdeg99$ and $2\fdeg23$. The root of both artifacts lies in a refraction process at the L3 chamfer on the cold side of the L3 (inside the cryostat). The \textit{L3 Angel Wings} occur through a pure refraction process by the L3 chamfer, while the \textit{Horseshoe} involves a subsequent reflection onto a mechanical edge of the L3 support frame (as shown in Fig.~\ref{horse_feature}). Both stray light paths follow the nominal optical path, but they are refracted by the L3 chamfer due to their off-axis coordinates. The features appear around the entire outer edge of the LSSTCam FoV and have comparable geometries, although with different spatial extents. Their prominence strongly depends on the magnitude of the bright object: for very bright objects (mag $<$ 3), both features cover a significant fraction of the focal plane as shown in Fig.~\ref{horse_feature}, but for fainter objects, both have a negligible spatial extension and are primarily constrained to the vignetted portion of the outermost detectors.

The Rubin stray light team is studying the feasibility of installing an additional light baffle in front of the L3 (on the warm side), and a detailed analysis of the design development is reported by Pollek et al. (2026) \cite{Pollek2026}. There are a variety of challenges to be addressed, both in the optical and thermal design. The LSSTCam has wavefront sensors and guiders at the four corners of the FoV, fully outside the scientific FoV; these corners must not be vignetted by any baffle up to $3\fdeg9$ off-axis, thus leading to some leaks of stray light in correspondence with the four corners. The L3 warm side is also constantly blown with dry air to prevent condensation and dust deposition, and the potential L3 baffle must not alter this flow. The ray tracing simulations shown in Fig.~\ref{horse_feature} demonstrate a good knowledge and reproducibility of both the stray light features, as well as the effectiveness of the proposed L3 baffling solution.

\subsubsection{The M2 baffle scattered light}
\label{baffle_m2}

\begin{figure}[!h]
    \centering
    \includegraphics[width=1\linewidth]{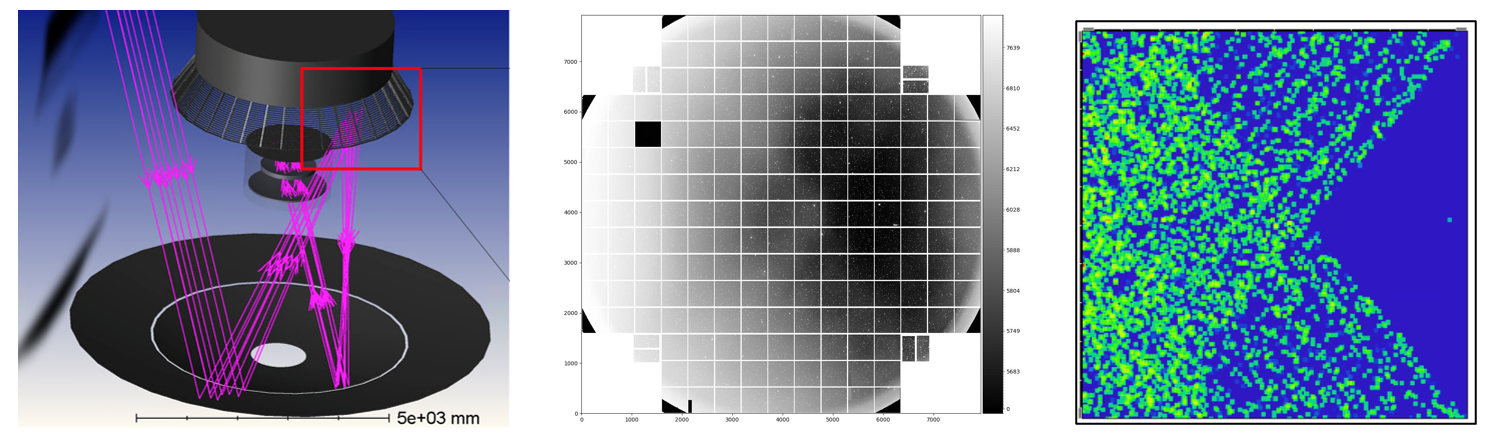}
    \caption{Left: stray light optical path generated by the M2 baffle, the rays reflected off M1, hit the M2 baffle, are scattered towards M3, and propagated to the LSSTCam focal plane. Centre: example on-sky of the light glow generated by the Moon with an illumination fraction of 25\% at 16$^\circ$ off-axis. Right: feature reproduction with the Ansys Zemax OpticStudio\registered non-sequential model.}
    \label{m2_baff_feature}
\end{figure}

The M2 baffle is assembled from a series of parallel light vane blades, as shown in Fig.~\ref{m2_baff_feature}.  The M2 baffle blades are made of black-anodized aluminum, which has a residual reflectance in the longer wavelength bands ($i$, $z$, $y$) that can reach up to 30\%. The observed stray light path comes from [M1/M3, M2 baffle, M3, LSSTCam], which occurs for off-axis angles between $12^\circ$ and $20^\circ$.  The scattered background light is visually evident only when triggered by the Moon, as can be seen in Fig.~\ref{m2_baff_feature}. A detailed analysis of the light scattered by the M2 baffle is reported by \cite{Taranto2026}, which compared the level of light background produced by the Moon and that measured with the CBP during in-dome testing. This stray light optical path will not be blocked by the LWS; however, the reflectance of the M2 baffle can be decreased if the surface is coated with Aeroglaze\registered Z306.

The effect is relevant only when the Moon is at an angular distance $<20^\circ$, while for the brightest stars (mag$_{z,y}<-2$) the amount of background light is at the level of a few electrons \cite{Taranto2026}. Low Surface Brightness science is the most affected by this problem. The expected mitigation strategy is to paint the M2 baffle with Aeroglaze\registered Z306 to decrease the residual reflectance by an order of magnitude. This will also decrease the stray light background in the image due to the higher scattering properties of the coating.  However, removal of the M2 baffle is a non-trivial engineering task that will need to wait for the right opportunity to arise.

\subsubsection{L2 Bananas}
\label{bananas}

\begin{figure}[!h]
    \centering
    \includegraphics[width=1\linewidth]{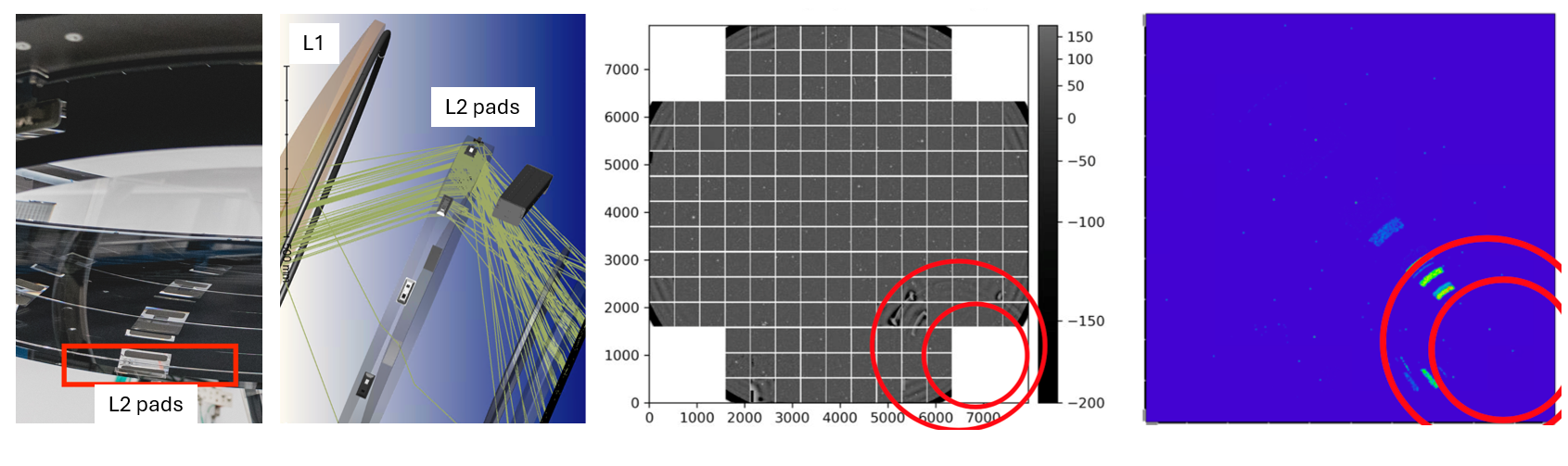}
    \caption{Left: close-up view of the L2 side support pad that can generate \textit{L2 Bananas} through the M1 direct path. Centre: sky-background-subtracted image with pairs of visible \textit{L2 Bananas} triggered by the Moon at $\sim25^\circ$ off-axis. Right: feature reproduction with the Ansys Zemax OpticStudio\registered non-sequential model.}
    \label{banana}
\end{figure}

The \textit{L2 Bananas} are a stray light artifact originating from a grazing reflection onto the L2 side support pads (Fig.~\ref{banana}). The \textit{L2 Bananas} stray light path originates from an M1 direct path (see Section \ref{fred}) that follows the trajectory [M1, L2 pads, L3, Focal Plane]. The feature is visible for off-axis coordinates between 23$\fdeg7$ and 26$^\circ$. The ray tracing simulations suggest that \textit{L2 Bananas} might also originate through the M3 direct path, but on-sky observations are not fully conclusive in this case.

The on-sky visibility window for the \textit{L2 Bananas} stray light path is expected to be closed by the completion of the installation of the LWS, which will have an angular acceptance of $\sim19\fdeg8$ x $\sim19\fdeg8$. The \textit{L2 Bananas} could still appear in LSST images if any artificial light source inside the dome illuminates the L2 support pads from the relevant off-axis range. Simulations indicate that a potential solution to remove these residual artifacts would come from a further extension of the mid-level baffle deployed for the removal of the scratched tape feature.

\subsubsection{Brush stroke}
\label{brush}

\begin{figure}[!h]
    \centering
    \includegraphics[width=1\linewidth]{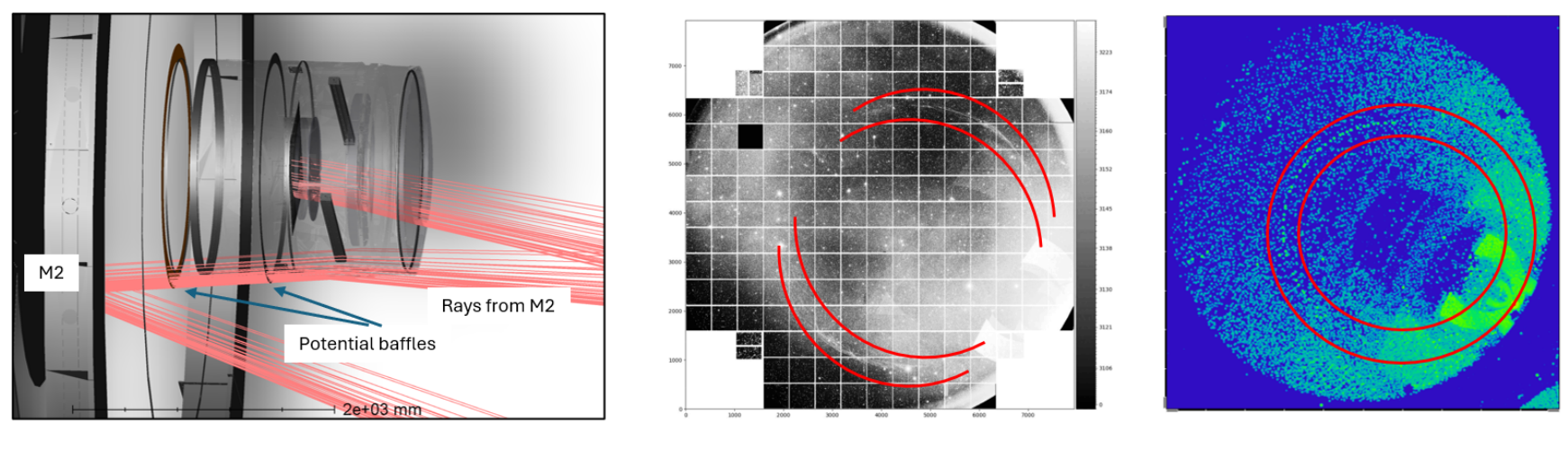}
    \caption{Left: Stray light optical path for the {\it brush stroke} feature, the rays reflected off the M1 and M2 and hit with a grazing angle the outer body of the LSSTCam before being reflected off the M3 and reaching the LSSTCam focal plane. Centre: Example on-sky of the feature highlighted in red generated by Canopus located at $1\fdeg9$ off-axis. Right: Feature reproduction with the Ansys Zemax OpticStudio\registered non-sequential model.}
    \label{brush_feature}
\end{figure}

The \textit{brush stroke} feature derives from the stray light path: [M1, M2, LSSTCam outer body,  M3, LSSTCam]. As shown in Fig.~\ref{brush_feature}, the stray light is due to a grazing-angle reflection onto the LSSTCam outer body of the photons traveling between M1 and M2 for off-axis angles between $1\fdeg3$ and $2\fdeg4$. This feature is visible only for bright objects (mag $<$ 2), and it becomes larger for increasing off-axis ranges. There are two baffling solutions under study: i) a series of small protruding annuli organized in a comb of baffles attached to the external sides of the LSSTCam; ii) covering the LSSTCam outer body with a sheet of fabric/rubber black material with good scattering properties. 
Both solutions will leave the telescope collecting area unaffected, with no additional vignetting factors.

\section{A system perspective to stray light}
\label{se_stray}

Table \ref{tab:rates} reports a preliminary census of the stray light artifacts observed in the LSSTCam image frames in the past year of observations (April 2025 - May 2026) during the commissioning and system optimization phases. For each feature, we applied the techniques described in Sections~\ref{emp_obs} and \ref{ray_trace} to infer the potential source/offender responsible for the stray light artifact, the envisioned mitigation strategy, the occurrence rate over $\sim$1 year, and the fractional area impacted on the LSSTCam focal plane. For $\sim$70\% of the features, we have already identified and confirmed their sources, and for $\sim$80\% of the artifacts, we have identified a possible mitigation strategy. The occurrence rate is based on a visual inspection and classification of the features visible in the image frames; for some features, the fractional area is a rough estimate because their extent varies across different image frames. The average fractional area can be compared to the maximum percentage of pixel loss at the LSSTCam focal plane ($<$4\% by the end of the 10-year survey) and the minimum area covered by science-grade imaging devices ($>$ 85\% over a 3.5-degree FoV) \cite{oss_doc}.

\begin{table}[t!]
\centering
\vspace{2em}
\begin{tabular}{l c c c c}
\hline
Artifact & Source & Mitigation & Occurrence Rate  & Fractional Area \\
\hline
\hline
Scratched Tape & M1 direct path & Mid-Level Baffle Extension & 7.39\% & 2.2\% * \\
Smudge & M1 direct path & Mid-Level Baffle Extension & 0.19\% & 0.28\% \\
Muddy Shoe & M1 direct path & Mid-Level Baffle Extension & 0.15\% & 2\%* \\
Pillow & Unknown & LWS installation & 0.26\% & 9\%* \\
Nibbled Disc & Unknown  & Mid-Level Baffle Extension & 0.10\% & 1.06 - 10\%* \\
L2 Bananas & L2 support pads & LWS installation & 0.73\% & 0.7\% \\
Rivet Holes & Holes in Top-end Baffle & Sealing of holes & 0.45\% & -- \\
Sharp Arcs & L2 edges \dag & TBD & TBD & 1\% $\Upsilon$ \\
Submarine & L1/L2 ghosts \dag & LWS installation (TBC) & 0.7\% $\Upsilon$ & 8.5\%* \\
L3 Horseshoe & L3 chamfer \& mount & Additional L3 edge baffle & 0.09\% & 8.8\% \\
\hline \\[-1em]
All artifacts & --- & --- & 10.00\% & --- \\
\multicolumn{2}{c}{Post Mid-Level Baffle Extension installation} & --- & 2.6\% & ---\\ 
\hline
\end{tabular}
\vspace{1em}
\caption{Occurrence rate of stray and scattered light artifacts that were prominent enough to be identified in visual inspection of full focal plane mosaic images. Rates are estimated from visual inspection of $\sim$116,000 images and are not expected to be complete or pure. The impacted area is estimated using the average size of the feature. The most prevalent artifact, the {\it scratched tape}, was effectively mitigated through installation of the mid-level baffle extension \citep{wagner2026}. \\ 
*\hspace{0.1cm}Fractional area of the artifact is variable; \dag \hspace{0.1cm} artifact origin to be confirmed; $\Upsilon$ preliminary estimate.}  
\label{tab:rates}
\end{table}

\begin{figure}[!h]
    \centering
    \includegraphics[width=0.85\linewidth]{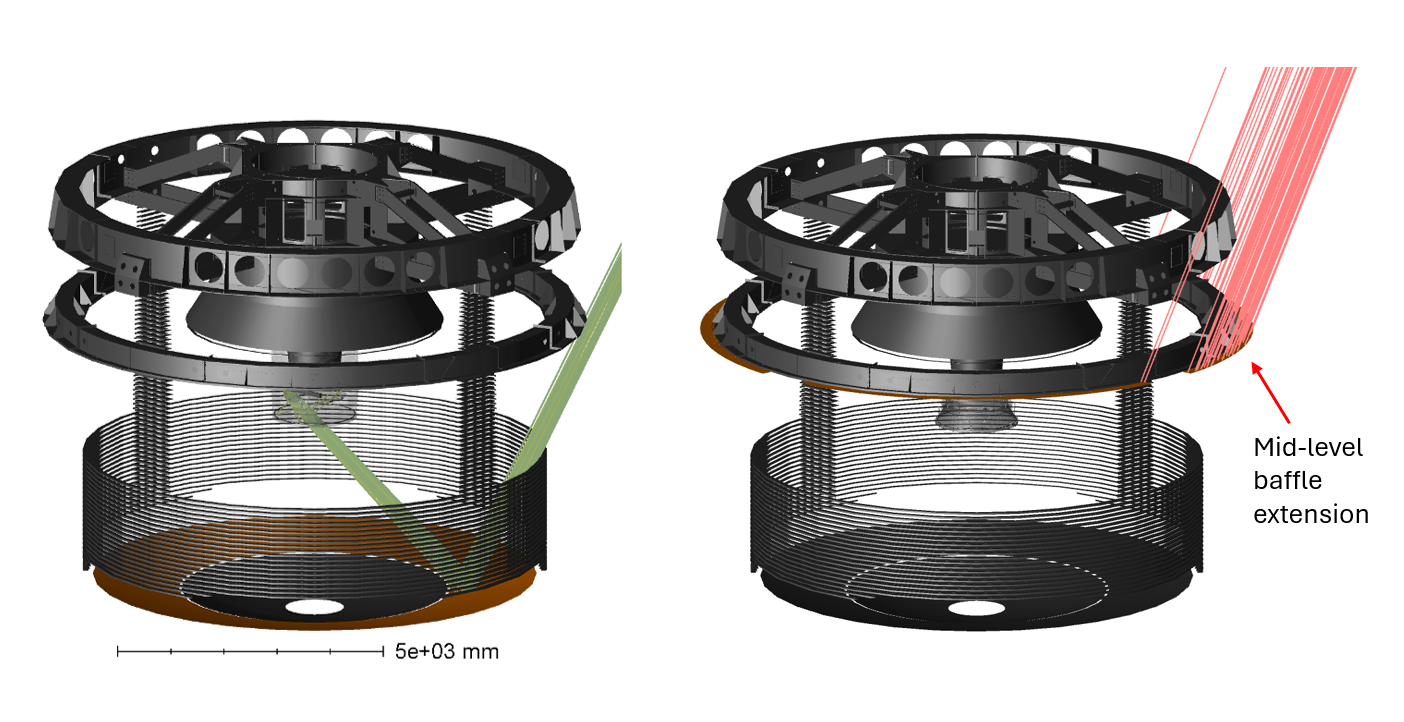}
    \caption{The mid-level baffle extension, in orange, (right) led to the dampening of the most prominent stray light feature (\textit{scratched tape}) that comes from light reaching the LSSTCam focal plane through an M1 direct path (left).}
    \label{mid-baf}
\end{figure}

The first stray light study carried out for Rubin \cite{ellis09,fred_06} in 2006 identified the main stray light paths for the telescope and camera and led to the issuing of $\sim$50 requirements related to the stray light prevention and mitigation. The requirements covered a wide range of design aspects to ensure that i) the optical windows are wedged to avoid straight ghost propagation, ii) the optical surfaces are coated with an anti-reflection coating to minimize ghost production, iii) the surfaces adjacent to the optical path that can potentially reflect and scatter stray light are black-coated, iv) the surface finishing and the blackening of the light baffles have low reflectivity towards the inside of the dome and high reflectivity towards the sky. The Rubin dome interior reflectivity is set to maximum $\sim$10\%, the louvers for dome ventilation are black and equipped with light traps. In addition, all dome openings need to provide scattering from at least two surfaces before transferring light into the dome environment; the dome design mitigates stray light towards the inside to facilitate daytime calibration activity. 
Despite an accurate optical analysis and the subsequent requirements flow-down at the system and subsystem level, prominent stray light contamination initially appeared in $\sim$10\% of image frames. This percentage was reduced to $\sim$2.6\% following the installation of the mid-level baffle extension (Fig.~\ref{mid-baf}) in early 2026 \cite{wagner2026}.

The stray light test campaign, deployed after the discovery of the stray light with ComCam, encompassed 38 test cases, for a total of $\sim$14 hours of testing on-sky and $\sim$19 hours of in-dome testing. The tests were designed to infer and/or confirm the origin of specific stray light artifacts, most of which were initially observed serendipitously during engineering time and scientific observations on-sky. Looking backward, the main limitation of the initial study was probably embedded in the intrinsic nature of the PST parameter used to quantify the level of stray light; the PST provides no information about the distribution of light on the detector focal plane, which is equivalent to having a single detector pixel covering the whole focal plane. The most critical stray light artifacts are those coming as structured features on the focal plane rather than diffused, scattered light background. Despite the effort of the analytical study and the systematic tracking of the requirements and their verification \cite{Selvy2018}, Rubin experienced a higher-than-expected amount of stray light during commissioning. Most of these features can be traced back to the incomplete status of the LWS. Some artifacts were expected from initial simulations, and their existence was confirmed empirically through on-sky testing, while new features were discovered serendipitously during on-sky observing.  Complex scientific systems like Rubin can require additional optimization and troubleshooting despite the adoption of rigorous construction requirements and quality assurance methodologies. Every new facility brings new challenges, which require creative ideas and solutions. Rubin has already benefited from the flexibility to implement new baffling solutions \cite{wagner2026}, and additional baffling solutions are currently being studied to further reduce the occurrence of stray light \cite{Pollek2026, Taranto2026}. To avoid stray light rather than flag and cut affected data, we are also exploring machine-learning classifiers (e.g., gradient-boosted decision trees) that could predict whether a given pointing will produce significant stray-light features, providing an additional weighting input to the Feature-Based Scheduler as labeled training data accumulates during pre-survey operations.

\subsection{Acknowledgments}

The corresponding author is grateful to Dr.\ Paolo Conconi and Dr.\ Andrea Bianco for the fruitful discussions that improved the initial modeling of the stray light features reported in this work. This material is based upon work supported in part by the National Science Foundation through Cooperative Agreements AST-1258333 and AST-2241526 and Cooperative Support Agreements AST-1202910 and 2211468 managed by the Association of Universities for Research in Astronomy (AURA), and the Department of Energy under Contract No. DE-AC02-76SF00515 with the SLAC National Accelerator Laboratory managed by Stanford University. Additional Rubin Observatory funding comes from private donations, grants to universities, and in-kind support from LSST-DA Institutional Members.
The authors acknowledge the contribution of the USC A of the INAF Scientific Directorate to the program "Italian participation in the Rubin LSST project".
The work of the authors was partially supported by the National Science Foundation Astronomy and Astrophysics Research Grants program through awards AST-2006340, AST-2307126, and AST-2407526. 

\bibliography{report} 

@ARTICLE{2026arXiv260323786V,
       author = {{Vera C Rubin Observatory Team} and {Acero Cuellar}, Tatiana and {Acosta}, Emily and {Adair}, Christina L and {Adari}, Prakruth and {Adelman McCarthy}, Jennifer K and {Alexov}, Anastasia and {Allbery}, Russ and {Allsman}, Robyn and {AlSayyad}, Yusra and {Amado}, Jhonatan and {Amouroux}, Nathan and {Antilogus}, Pierre and {Aracena Alcayaga}, Alexis and {Aravena Rojas}, Gonzalo and {Araya Cortes}, Claudio H and {Aubourg}, Eric and {Axelrod}, Tim S and {Banovetz}, John and {Barria}, Carlos and {E Bauer}, Amanda and {Bauman}, Brian J and {Bechtol}, Ellen and {Bechtol}, Keith and {Becker}, Andrew C and {Becker}, Valerie R and {Beckett}, Mark G and {Bellm}, Eric C and {Bernardinelli}, Pedro H and {Bettina Bianco}, Federica and {Blum}, Robert D and {Bogart}, Joanne and {Bolton}, Adam and {Booth}, Michael T and {Bosch}, James F and {Boucaud}, Alexandre and {Boutigny}, Dominique and {Bovill}, Robert A and {Bradshaw}, Andrew and {Bregeon}, Johan and {Brescia}, Massimo and {Brondel}, Brian J and {Broughton}, Alexander and {Budlong}, Audrey and {Buffat}, Dimitri and {Canestrari}, Rodolfo and {Caplar}, Neven and {Carlin}, Jeffrey L and {Ceballo}, Ross and {Chandler}, Colin Orion and {Chang}, Chihway and {Emerson}, Glenaver Charles and {Chiang}, Hsin Fang and {Chiang}, James and {Choi}, Yumi and {Christensen}, Eric J and {Claver}, Charles F and {Clements}, Andy W and {Cockrum}, Joseph J and {Cohen Tanugi}, Johann and {Colleoni}, Franco and {Combet}, Celine and {Connolly}, Andrew J and {Constanzo Cordova}, Julio Eduardo and {E Contreras}, Hans and {Crenshaw}, John Franklin and {Dagoret Campagne}, Sylvie and {Daniel}, Scott F and {Daruich}, Felipe and {Daubard}, Guillaume and {Daues}, Greg and {Dennihy}, Erik and {Deppe}, Stephanie J H and {Digel}, Seth W and {E Doherty}, Peter and {Doux}, Cyrille and {Drlica Wagner}, Alex and {Dubois Felsmann}, Gregory P and {Economou}, Frossie and {Eiger}, Orion and {Eisert}, Lukas and {Eisner}, Alan M and {Englert}, Anthony and {Erb}, Baden and {Fabrega}, Juan A and {Fagrelius}, Parker and {Fanning}, Kevin and {Fausti Neto}, Angelo and {Ferguson}, Peter S and {Ferte}, Agnes and {Findeisen}, Krzysztof and {Levine}, Merlin Fisher and {Fonseca Alvarez}, Gloria and {Foss}, Michael D and {Fouchez}, Dominique and {Fuchs}, Dan C and {Fu}, Shenming and {Gangler}, Emmanuel and {Gaponenko}, Igor and {Garcia}, Julen and {Gates}, John H and {Gill}, Ranpal K and {Giro}, Enrico and {Glanzman}, Thomas and {Godoy}, Robinson and {Goodenow}, Iain and {Gorsuch}, Miranda R and {Gower}, Michelle and {Graham}, Melissa L and {Granvik}, Mikael and {Greenstreet}, Sarah and {Guan}, Wen and {Guillemin}, Thibault and {Guy}, Leanne P and {Hascall}, Diane and {Hascall}, Patrick A and {Heinze}, Aren Nathaniel and {Hernandez}, Fabio and {Herner}, Kenneth and {Herrold}, Ardis and {Higgs}, Clare R and {Hoblitt}, Joshua and {Howard}, Erin Leigh and {Hyun}, Minhee and {Ibsen}, Amanda and {Ingraham}, Patrick and {Irving}, David H and {Ivezic}, Zeljko and {Jacoby}, Suzanne H and {Jannuzi}, Buell T and {Jarugula}, Sreevani and {Jee}, M James and {Jenness}, Tim and {Jennings}, Toby C and {Jeremie}, Andrea and {Jernigan}, Garrett and {Jimenez Mejias}, David and {Johnson}, Anthony S and {Jones}, R Lynne and {Jones}, Roger William Lewis and {Juramy Gilles}, Claire and {Juric}, Mario and {Kahn}, Steven M and {Bryce Kalmbach}, J and {Kang}, Yijung and {Kannawadi}, Arun and {Kantor}, Jeffrey P and {Karavakis}, Edward and {Kelkar}, Kshitija and {Kelvin}, Lee S and {Kleinman}, Scot J and {Kotov}, Ivan V and {Kovacs}, Gabor and {Kowalik}, Mikolaj and {Krabbendam}, Victor L and {Krughoff}, K Simon and {Kubanek}, Petr and {Kurlander}, Jacob A and {Kusulja}, Mile and {Lage}, Craig S and {Lago}, Paulo J A and {Laliotis}, Katherine and {Lange}, Travis and {Laporte}, Didier and {Lau}, Ryan M and {Lazarte}, Juan Carlos and {Le Boulch}, Quentin and {Leget}, Pierre Francois and {Le Guillou}, Laurent and {Levine}, Benjamin and {Liang}, Ming and {Liang}, Shuang and {Lim}, Kian Tat and {von der Linden}, Anja and {Lin}, Huan and {Lopez}, Margaux and {Lopez Toro}, Juan J and {Love}, Peter and {Lupton}, Robert H and {Lust}, Nate B and {MacArthur}, Lauren A and {MacBride}, Sean Patrick and {Madejski}, Greg M and {Mainetti}, Gabriele and {Margheim}, Steven J and {Markiewicz}, Thomas W and {Marshall}, Phil and {Marshall}, Stuart and {Maulen}, Guido and {Mau}, Sidney and {May}, Morgan and {McCormick}, Jeremy and {McKay}, David and {McKercher}, Robert and {Megias Homar}, Guillem and {Meisner}, Aaron M and {Menanteau}, Felipe and {Mentzer}, Heather R and {Metzger}, Kristen and {E Meyers}, Joshua},
        title = "{The Vera C. Rubin Observatory Data Preview 1}",
      journal = {arXiv e-prints},
         year = 2026,
        month = mar,
          eid = {arXiv:2603.23786},
        pages = {arXiv:2603.23786},
          doi = {10.48550/arXiv.2603.23786},
archivePrefix = {arXiv},
       eprint = {2603.23786},
 primaryClass = {astro-ph.IM},
       adsurl = {https://ui.adsabs.harvard.edu/abs/2026arXiv260323786V}
}

@inproceedings{Stalder20,
author = {Brian Stalder and Kevin Reil and Chuck Claver and Ming Liang and Tei Wei Tsai and Travis Lange and Justine Haupt and Oliver Wiecha and Margaux Lopez and Gary Poczulp and Diane Hascall and Douglas Neill and Jacques Sebag and Brian Johnson and Neill Mills and Myung Cho and Homer Neal and Scott Newbry and Shawn Osier and Rafe Schindler and Dmitry Onoprienko and Anthony Johnson and R. Glenn Morris and Max Turri and Alan Eisner and Stephen Cisneros and Van Xiong and Michael Huffer and Gregg Thayer and Ronald Harris and Anthony Borstad and Anthony Tache and William Schoening and J. Anthony Tyson and Craig Lage and Merlin Fisher-Levine and Robert Lupton and Andres Plazas and Felipe Menanteau and Htut Khine Htay Win and Stephen Pietrowicz and James Howard},
title = {{Rubin commissioning camera: integration, functional testing, and lab performance}},
volume = {11447},
booktitle = {Ground-based and Airborne Instrumentation for Astronomy VIII},
editor = {Christopher J. Evans and Julia J. Bryant and Kentaro Motohara},
organization = {International Society for Optics and Photonics},
publisher = {SPIE},
pages = {114470L},
year = {2020},
doi = {10.1117/12.2561132},
URL = {https://doi.org/10.1117/12.2561132}
}

@misc{sitcomtn149,
  title = {SITCOMTN-149: An Interim Report on the LSSTComCam On-Sky Campaign}, 
  author = {{Rubin Observatory Project}},
  howpublished = {\url{https://sitcomtn-149.lsst.io/SITCOMTN-149.pdf}},
}

@article{Bock_2026,
doi = {10.3847/1538-4357/ae2be2},
url = {https://doi.org/10.3847/1538-4357/ae2be2},
year = {2026},
month = {feb},
publisher = {The American Astronomical Society},
volume = {999},
number = {1},
pages = {139},
author = {Bock, James J. and Aboobaker, Asad M. and Adamo, Joseph and Akeson, Rachel and Alred, John M. and Alibay, Farah and Ashby, Matthew L. N. and Bach, Yoonsoo P. and Bleem, Lindsey E. and Bolton, Douglas and Braun, David F. and Bruton, Sean and Bryan, Sean A. and Chang, Tzu-Ching and Chen, Shuang-Shuang and Cheng, Yun-Ting and Cheshire, James R. and Chiang, Yi-Kuan and de Janvry, Jean Choppin and Condon, Samuel and Cook, Walter R. and Cooray, Asantha and Crill, Brendan P. and Cukierman, Ari J. and Doré, Olivier and Dowell, C. Darren and Dubois-Felsmann, Gregory P. and Eifler, Tim and Everett, Spencer and Fabinsky, Beth E. and Faisst, Andreas L. and Fanson, James L. and Farrington, Allen H. and Fatahi, Tamim and Fazar, Candice M. and Feder, Richard M. and Frater, Eric H. and Grasshorn Gebhardt, Henry S. and Giri, Utkarsh and Goldina, Tatiana and Gorjian, Varoujan and Habib, Salman and Hart, William G. and Heinrich, Chen and Hora, Joseph L. and Huai, Zhaoyu and Hui, Howard and Jo, Young-Soo and Jeong, Woong-Seob and Kang, Jae Hwan and Kang, Miju and Kecman, Branislav and Kim, Chul-Hwan and Kim, Jaeyeong and Kim, Minjin and Kim, Young-Jun and Kim, Yongjung and Kirkpatrick, J. Davy and Kobayashi, Yosuke and Korngut, Phil M. and Krause, Elisabeth and Lee, Bomee and Lee, Ho-Gyu and Lee, Jae-Joon and Lee, Jeong-Eun and Lisse, Carey M. and Mariani, Giacomo and Masters, Daniel C. and Mauskopf, Philip D. and Melnick, Gary J. and Minasyan, Mary H. and Mirocha, Jordan and Miyasaka, Hiromasa and Moore, Anne and Moore, Bradley D. and Murgia, Giulia and Naylor, Bret J. and Nelson, Christina and Nguyen, Chi H. and Nguyen, Hien T. and Noh, Jinyoung K. and Padin, Stephen and Paladini, Roberta and Park, Sung-Joon and Penanen, Konstantin I. and Putnam, Dustin S. and Pyo, Jeonghyun and Ramachandra, Nesar and Ramanathan, Keshav and Rustamkulov, Zafar and Reiley, Daniel J. and Rice, Eric B. and Rocca, Jennifer M. and Seok, Ji Yeon and Smith, Roger and Stober, Jeremy and Susca, Sara and Teplitz, Harry I. and Thelen, Michael P. and Tolls, Volker and Torrini, Gabriela and Trangsrud, Amy R. and Unwin, Stephen and Velicheti, Phani and Wang, Pao-Yu and Wen, Robin Y. and Werner, Michael W. and Williams, Abby E. and Williamson, Ross and Wincentsen, James and Windhorst, Rogier A. and Yang, Soung-Chul and Yang, Yujin and Zemcov, Michael},
title = {The SPHEREx Satellite Mission},
journal = {The Astrophysical Journal}
}

@misc{esa2023,
  title        = {Seeking Euclid's hidden stars: commissioning looks up},
  author       = {},
  year         = 2023,
  note         = {\url{https://www.esa.int/Science_Exploration/Space_Science/Euclid/Seeking_Euclid_s_hidden_stars_commissioning_looks_up}}
}

@misc{vst2020,
  title        = {Very Large Telescope Paranal Science Operations OmegaCAM User Manual},
  author       = {},
  year         = 2020,
  note         = {\url{https://www.eso.org/sci/facilities/paranal/instruments/omegacam/doc/VST-MAN-OCM-23100-3110_p106_v2.pdf}}
}

@misc{fred_06,
    author = {Scott Ellis},
    title = {{Stray Light Analysis of the LSST 8.4m Wide Field Telescope}},
    howpublished = "Photon Engineering, LLC 440 S. Williams Boulevard, Suite 106 Tucson, Arizona 85711 14 November 2006"}

@article{Flaugher_2015,
doi = {10.1088/0004-6256/150/5/150},
url = {https://doi.org/10.1088/0004-6256/150/5/150},
year = {2015},
month = {oct},
publisher = {The American Astronomical Society},
volume = {150},
number = {5},
pages = {150},
author = {Flaugher, B. and Diehl, H. T. and Honscheid, K. and Abbott, T. M. C. and Alvarez, O. and Angstadt, R. and Annis, J. T. and Antonik, M. and Ballester, O. and Beaufore, L. and Bernstein, G. M. and Bernstein, R. A. and Bigelow, B. and Bonati, M. and Boprie, D. and Brooks, D. and Buckley-Geer, E. J. and Campa, J. and Cardiel-Sas, L. and Castander, F. J. and Castilla, J. and Cease, H. and Cela-Ruiz, J. M. and Chappa, S. and Chi, E. and Cooper, C. and da Costa, L. N. and Dede, E. and Derylo, G. and DePoy, D. L. and de Vicente, J. and Doel, P. and Drlica-Wagner, A. and Eiting, J. and Elliott, A. E. and Emes, J. and Estrada, J. and Fausti Neto, A. and Finley, D. A. and Flores, R. and Frieman, J. and Gerdes, D. and Gladders, M. D. and Gregory, B. and Gutierrez, G. R. and Hao, J. and Holland, S. E. and Holm, S. and Huffman, D. and Jackson, C. and James, D. J. and Jonas, M. and Karcher, A. and Karliner, I. and Kent, S. and Kessler, R. and Kozlovsky, M. and Kron, R. G. and Kubik, D. and Kuehn, K. and Kuhlmann, S. and Kuk, K. and Lahav, O. and Lathrop, A. and Lee, J. and Levi, M. E. and Lewis, P. and Li, T. S. and Mandrichenko, I. and Marshall, J. L. and Martinez, G. and Merritt, K. W. and Miquel, R. and Muñoz, F. and Neilsen, E. H. and Nichol, R. C. and Nord, B. and Ogando, R. and Olsen, J. and Palaio, N. and Patton, K. and Peoples, J. and Plazas, A. A. and Rauch, J. and Reil, K. and Rheault, J.-P. and Roe, N. A. and Rogers, H. and Roodman, A. and Sanchez, E. and Scarpine, V. and Schindler, R. H. and Schmidt, R. and Schmitt, R. and Schubnell, M. and Schultz, K. and Schurter, P. and Scott, L. and Serrano, S. and Shaw, T. M. and Smith, R. C. and Soares-Santos, M. and Stefanik, A. and Stuermer, W. and Suchyta, E. and Sypniewski, A. and Tarle, G. and Thaler, J. and Tighe, R. and Tran, C. and Tucker, D. and Walker, A. R. and Wang, G. and Watson, M. and Weaverdyck, C. and Wester, W. and Woods, R. and Yanny, B. and (The DES Collaboration)},
title = {THE DARK ENERGY CAMERA},
journal = {The Astronomical Journal}
}

@ARTICLE{Ivezic19,
       author = {{Ivezi{\'c}}, {\v{Z}}eljko and {Kahn}, Steven M. and {Tyson}, J. Anthony and {Abel}, Bob and {Acosta}, Emily and {Allsman}, Robyn and {Alonso}, David and {AlSayyad}, Yusra and {Anderson}, Scott F. and {Andrew}, John and {Angel}, James Roger P. and {Angeli}, George Z. and {Ansari}, Reza and {Antilogus}, Pierre and {Araujo}, Constanza and {Armstrong}, Robert and {Arndt}, Kirk T. and {Astier}, Pierre and {Aubourg}, {\'E}ric and {Auza}, Nicole and {Axelrod}, Tim S. and {Bard}, Deborah J. and {Barr}, Jeff D. and {Barrau}, Aurelian and {Bartlett}, James G. and {Bauer}, Amanda E. and {Bauman}, Brian J. and {Baumont}, Sylvain and {Bechtol}, Ellen and {Bechtol}, Keith and {Becker}, Andrew C. and {Becla}, Jacek and {Beldica}, Cristina and {Bellavia}, Steve and {Bianco}, Federica B. and {Biswas}, Rahul and {Blanc}, Guillaume and {Blazek}, Jonathan and {Blandford}, Roger D. and {Bloom}, Josh S. and {Bogart}, Joanne and {Bond}, Tim W. and {Booth}, Michael T. and {Borgland}, Anders W. and {Borne}, Kirk and {Bosch}, James F. and {Boutigny}, Dominique and {Brackett}, Craig A. and {Bradshaw}, Andrew and {Brandt}, William Nielsen and {Brown}, Michael E. and {Bullock}, James S. and {Burchat}, Patricia and {Burke}, David L. and {Cagnoli}, Gianpietro and {Calabrese}, Daniel and {Callahan}, Shawn and {Callen}, Alice L. and {Carlin}, Jeffrey L. and {Carlson}, Erin L. and {Chandrasekharan}, Srinivasan and {Charles-Emerson}, Glenaver and {Chesley}, Steve and {Cheu}, Elliott C. and {Chiang}, Hsin-Fang and {Chiang}, James and {Chirino}, Carol and {Chow}, Derek and {Ciardi}, David R. and {Claver}, Charles F. and {Cohen-Tanugi}, Johann and {Cockrum}, Joseph J. and {Coles}, Rebecca and {Connolly}, Andrew J. and {Cook}, Kem H. and {Cooray}, Asantha and {Covey}, Kevin R. and {Cribbs}, Chris and {Cui}, Wei and {Cutri}, Roc and {Daly}, Philip N. and {Daniel}, Scott F. and {Daruich}, Felipe and {Daubard}, Guillaume and {Daues}, Greg and {Dawson}, William and {Delgado}, Francisco and {Dellapenna}, Alfred and {de Peyster}, Robert and {de Val-Borro}, Miguel and {Digel}, Seth W. and {Doherty}, Peter and {Dubois}, Richard and {Dubois-Felsmann}, Gregory P. and {Durech}, Josef and {Economou}, Frossie and {Eifler}, Tim and {Eracleous}, Michael and {Emmons}, Benjamin L. and {Fausti Neto}, Angelo and {Ferguson}, Henry and {Figueroa}, Enrique and {Fisher-Levine}, Merlin and {Focke}, Warren and {Foss}, Michael D. and {Frank}, James and {Freemon}, Michael D. and {Gangler}, Emmanuel and {Gawiser}, Eric and {Geary}, John C. and {Gee}, Perry and {Geha}, Marla and {Gessner}, Charles J.~B. and {Gibson}, Robert R. and {Gilmore}, D. Kirk and {Glanzman}, Thomas and {Glick}, William and {Goldina}, Tatiana and {Goldstein}, Daniel A. and {Goodenow}, Iain and {Graham}, Melissa L. and {Gressler}, William J. and {Gris}, Philippe and {Guy}, Leanne P. and {Guyonnet}, Augustin and {Haller}, Gunther and {Harris}, Ron and {Hascall}, Patrick A. and {Haupt}, Justine and {Hernandez}, Fabio and {Herrmann}, Sven and {Hileman}, Edward and {Hoblitt}, Joshua and {Hodgson}, John A. and {Hogan}, Craig and {Howard}, James D. and {Huang}, Dajun and {Huffer}, Michael E. and {Ingraham}, Patrick and {Innes}, Walter R. and {Jacoby}, Suzanne H. and {Jain}, Bhuvnesh and {Jammes}, Fabrice and {Jee}, M. James and {Jenness}, Tim and {Jernigan}, Garrett and {Jevremovi{\'c}}, Darko and {Johns}, Kenneth and {Johnson}, Anthony S. and {Johnson}, Margaret W.~G. and {Jones}, R. Lynne and {Juramy-Gilles}, Claire and {Juri{\'c}}, Mario and {Kalirai}, Jason S. and {Kallivayalil}, Nitya J. and {Kalmbach}, Bryce and {Kantor}, Jeffrey P. and {Karst}, Pierre and {Kasliwal}, Mansi M. and {Kelly}, Heather and {Kessler}, Richard and {Kinnison}, Veronica and {Kirkby}, David and {Knox}, Lloyd and {Kotov}, Ivan V. and {Krabbendam}, Victor L. and {Krughoff}, K. Simon and {Kub{\'a}nek}, Petr and {Kuczewski}, John and {Kulkarni}, Shri and {Ku}, John and {Kurita}, Nadine R. and {Lage}, Craig S. and {Lambert}, Ron and {Lange}, Travis and {Langton}, J. Brian and {Le Guillou}, Laurent and {Levine}, Deborah and {Liang}, Ming and {Lim}, Kian-Tat and {Lintott}, Chris J. and {Long}, Kevin E. and {Lopez}, Margaux and {Lotz}, Paul J. and {Lupton}, Robert H. and {Lust}, Nate B. and {MacArthur}, Lauren A. and {Mahabal}, Ashish and {Mandelbaum}, Rachel and {Markiewicz}, Thomas W. and {Marsh}, Darren S. and {Marshall}, Philip J. and {Marshall}, Stuart and {May}, Morgan and {McKercher}, Robert and {McQueen}, Michelle and {Meyers}, Joshua and {Migliore}, Myriam and {Miller}, Michelle and {Mills}, David J.},
        title = "{LSST: From Science Drivers to Reference Design and Anticipated Data Products}",
      journal = {ApJ},
         year = 2019,
        month = mar,
       volume = {873},
       number = {2},
          eid = {111},
        pages = {111},
          doi = {10.3847/1538-4357/ab042c},
archivePrefix = {arXiv},
       eprint = {0805.2366},
 primaryClass = {astro-ph},
       adsurl = {https://ui.adsabs.harvard.edu/abs/2019ApJ...873..111I}
}

@unpublished{wagner2026,
author  = "{Drlica-Wagner}, Alex and {Taranto}, Alessio and {Rodeghiero}, Gabriele and {Meyers}, Josh E. and {Neill}, D. R. and others",
title   = "Study of a large-angle off-axis stray light path in Rubin Observatory commissioning",
note    = "Paper number 
14147-121, to be published on Proc. of SPIE 2026 for Ground-based and Airborne Telescopes XI"
}

@unpublished{Taranto2026,
author  = "{Taranto}, Alessio and {Rodeghiero}, Gabriele and {Rosignoli}, Luca and {Urbach}, K. E. and {M\"uller}, Fritz",
title   = "Modeling of the diffuse background produced by the Vera C. Rubin Observatory M2 baffle scattered light",
note    = "Paper number 
14152-98, to be published on Proc. of SPIE 2026 for Modeling, Systems Engineering, and Project Management for Astronomy XII"
}

@unpublished{Pollek2026,
author  = "{Pollek}, Hannah M. M. and {Rodeghiero}, Gabriele and {Andrew}, John and {Drlica-Wagner}, Alex and {Taranto}, Alessio",
title   = "Mechanical studies of an additional light baffle for the LSST camera",
note    = "Paper number 
14152-91, to be published on Proc. of SPIE 2026 for Modeling, Systems Engineering, and Project Management for Astronomy XII"
}

@unpublished{Pai2026,
author  = "{Pai}, Aashay and {Drlica-Wagner}, Alex and {Kelvin}, Lee S. and {Meyers}, Joshua E. and {Urbach}, Elana K. and Mueller, Fritz and {Lupton}, Robert H.",
title   = "Optical ghosts: quantifying their impact and using them to probe LSSTCam optics",
note    = "Paper number 
14152-92, to be published on Proc. of SPIE 2026 for Modeling, Systems Engineering, and Project Management for Astronomy XII"
}

@misc{batoid19,
title = {batoid},
author = {Meyers, Joshua E. and Kirkby, David and Thomas, David},
abstractNote = {Batoid is a c++ backed python optical raytracer. It uses geometric optics to characterize the optical performance of survey telescopes such as LSST, Subaru-HSC, Blanco-DECam, or Mayall-DESI.},
doi = {10.11578/dc.20200708.1},
url = {https://doi.org/10.11578/dc.20200708.1},
howpublished = {[Computer Software] \url{https://doi.org/10.11578/dc.20200708.1}},
year = {2019},
month = {oct}
}

@inproceedings{Selvy2018,
author = {Brian M. Selvy and Austin Roberts and Michael Reuter and Charles (Chuck) Claver and Gabriele Comoretto and Tim Jenness and Wil O'Mullane and Andrew Serio and Rob Bovill and Jacques Sebag and Sandrine Thomas and Manas Bajaj and Dirk Zwemer and Brian Van Klaveren},
title = {{V\&V planning and execution in an integrated model-based engineering environment using MagicDraw, Syndeia, and Jira}},
volume = {10705},
booktitle = {Modeling, Systems Engineering, and Project Management for Astronomy VIII},
editor = {George Z. Angeli and Philippe Dierickx},
organization = {International Society for Optics and Photonics},
publisher = {SPIE},
pages = {107050U},
year = {2018},
doi = {10.1117/12.2310125},
URL = {https://doi.org/10.1117/12.2310125}
}

@INPROCEEDINGS{Angel2000,
       author = {{Angel}, R. and {Lesser}, M. and {Sarlot}, R. and {Dunham}, E.},
        title = "{Design for an 8-m Telescope with a 3 Degree Field at f/1.25: The Dark Matter Telescope}",
    booktitle = {Imaging the Universe in Three Dimensions},
         year = 2000,
       editor = {{van Breugel}, W. and {Bland-Hawthorn}, J.},
       series = {Astronomical Society of the Pacific Conference Series},
       volume = {195},
        month = jan,
        pages = {81},
       adsurl = {https://ui.adsabs.harvard.edu/abs/2000ASPC..195...81A}
}

@INPROCEEDINGS{Tyson2001,
       author = {{Tyson}, A. and {Angel}, R.},
        title = "{The Large-aperture Synoptic Survey Telescope.}",
    booktitle = {The New Era of Wide Field Astronomy},
         year = 2001,
       editor = {{Clowes}, Roger and {Adamson}, Andrew and {Bromage}, Gordon},
       series = {Astronomical Society of the Pacific Conference Series},
       volume = {232},
        month = jan,
        pages = {347},
       adsurl = {https://ui.adsabs.harvard.edu/abs/2001ASPC..232..347T}
}

@INPROCEEDINGS{Tyson2002,
       author = {{Tyson}, Tony and {Wittman}, David and {Hennawi}, Joe and {Spergel}, David},
        title = "{LSST as a precision probe of dark energy}",
    booktitle = {APS April Meeting Abstracts},
         year = 2002,
       series = {APS Meeting Abstracts},
        month = apr,
          eid = {Y6.004},
        pages = {Y6.004},
       adsurl = {https://ui.adsabs.harvard.edu/abs/2002APS..APR.Y6004T}
}

@MISC{Mountain2018,
       author = {{Mountain}, Charles Mattias and {Blum}, Robert D},
        title = "{NSF's National Optical-infrared Astronomy Research Laboratory: Pre-Operations of the Vera C. Rubin Observatory}",
 howpublished = {NSF Award Number 1836783. Directorate for Mathematical and Physical Sciences, Division Of Astronomical Sciences. 2018.},
         year = 2018,
        month = oct,
        pages = {36783},
       adsurl = {https://ui.adsabs.harvard.edu/abs/2018nsf....1836783M}
}

@ARTICLE{Chang2021,
       author = {{Chang}, Chihway and {Drlica-Wagner}, Alex and {Kent}, Stephen M. and {Nord}, Brian and {Wang}, Donah Michelle and {Wang}, Michael H.~L.~S.},
        title = "{A Machine Learning Approach to the Detection of Ghosting and Scattered Light Artifacts in Dark Energy Survey Images}",
      journal = {arXiv e-prints},
         year = 2021,
        month = may,
          eid = {arXiv:2105.10524},
        pages = {arXiv:2105.10524},
archivePrefix = {arXiv},
       eprint = {2105.10524},
 primaryClass = {astro-ph.IM},
       adsurl = {https://ui.adsabs.harvard.edu/abs/2021arXiv210510524C}
}

@unpublished{Kent2013,
    author = {Kent, Stephen M.},
    title = "{Ghost Images in DECam}",
    collaboration  = "DES",
    year = {2013},
    note = "{FERMILAB-SLIDES-20-114-SCD}",
    doi = {10.2172/1690257},
}

@ARTICLE{Tanoglidis2022,
       author = {{Tanoglidis}, D. and {{\'C}iprijanovi{\'c}}, A. and {Drlica-Wagner}, A. and {Nord}, B. and {Wang}, M.~H.~L.~S. and {Amsellem}, A. Jacob and {Downey}, K. and {Jenkins}, S. and {Kafkes}, D. and {Zhang}, Z.},
        title = "{DeepGhostBusters: Using Mask R-CNN to detect and mask ghosting and scattered-light artifacts from optical survey images}",
      journal = {Astronomy and Computing},
         year = 2022,
        month = apr,
       volume = {39},
          eid = {100580},
        pages = {100580},
          doi = {10.1016/j.ascom.2022.100580},
archivePrefix = {arXiv},
       eprint = {2109.08246},
 primaryClass = {astro-ph.IM},
       adsurl = {https://ui.adsabs.harvard.edu/abs/2022A&C....3900580T}
}

@MISC{lsstcam_ref,
       author = {{SLAC National Accelerator Laboratory} and {NSF-DOE Vera C. Rubin Observatory}},
        title = "{The LSST Camera (LSSTCam)}",
         year = 2025,
        month = jan,
          doi = {10.71929/RUBIN/2571927},
    publisher = {SLAC National Accelerator Laboratory},
       adsurl = {https://ui.adsabs.harvard.edu/abs/2025rubn.inst...36S}
}

@ARTICLE{blum25,
       author = {{Blum}, B.},
        title = "{Rubin Observatory: It Is Happening!}",
      journal = {The NOIRLab Mirror},
         year = 2025,
        month = jan,
       volume = {8},
        pages = {22},
       adsurl = {https://ui.adsabs.harvard.edu/abs/2025Mirro...8...22B}
}

@unpublished{Aranda2026,
author  = "{Aranda Sanchez}, Sebastain",
title   = "What is LOVE: learnings and challenges in developing a real time monitoring system for the Vera C Rubin Observatory operations",
note    = "Paper number 
14155-155, to be published on Proc. of SPIE 2026 for Software and Cyberinfrastructure for Astronomy IX"
}

@article{Liu_2024,
doi = {10.3847/1538-4357/ad3635},
url = {https://doi.org/10.3847/1538-4357/ad3635},
year = {2024},
month = {may},
publisher = {The American Astronomical Society},
volume = {967},
number = {1},
pages = {10},
author = {Liu, S. and Wood-Vasey, W. M. and Armstrong, R. and Narayan, G. and SÃ¡nchez, B. O. and The Dark Energy Science Collaboration},
title = {Testing the LSST Difference Image Analysis Pipeline Using Synthetic Source Injection Analysis},
journal = {The Astrophysical Journal},
}

@misc{stellarium,
  title = {Stellarium 26.1 User Guide}, 
  howpublished ={\url{https://stellarium.org/files/guide.pdf}},
}

@article{Mondrik_2023,
doi = {10.1088/1538-3873/acbe1c},
url = {https://doi.org/10.1088/1538-3873/acbe1c},
year = {2023},
month = {mar},
publisher = {The Astronomical Society of the Pacific},
volume = {135},
number = {1045},
pages = {035001},
author = {Mondrik, Nicholas and Coughlin, Michael and Betoule, Marc and Bongard, SÃ©bastien and Rice, Joseph P. and Shaw, Ping-Shine and Stubbs, Christopher W. and Woodward, John T. and LSST Dark Energy Science Collaboration},
title = {Measurement of Telescope Transmission Using a Collimated Beam Projector},
journal = {Publications of the Astronomical Society of the Pacific},
}

@misc{string_zem,
  title = {Identifying specific rays using filter strings}, 
  howpublished ={\url{https://optics.ansys.com/hc/en-us/articles/42661949492755-Identifying-specific-rays-using-filter-strings}},
}

@inproceedings{ellis09,
author = {K. Scott Ellis},
title = {{Stray light characteristics of the Large Synoptic Survey Telescope (LSST)}},
volume = {7427},
booktitle = {Optical Modeling and Performance Predictions IV},
editor = {Mark A. Kahan},
organization = {International Society for Optics and Photonics},
publisher = {SPIE},
pages = {742708},
year = {2009},
doi = {10.1117/12.830599},
URL = {https://doi.org/10.1117/12.830599}
}

@proceeding{sebag16, author = {Sebag, J. and Gressler, W. and Liang, M. and Neill, D. and Araujo-Hauck, C. and Andrew, J. and Angeli, G. and Cho, M. and Claver, C. and Daruich, F. and Gessner, C. and Hileman, E. and Krabbendam, V. and Muller, G. and Poczulp, G. and Repp, R. and Wiecha, O. and Xin, B. and Kenagy, K. and Martin, H. M. and Tuell, M. T. and West, S. C.}, title = { LSST primary/tertiary monolithic mirror }, journal = {Proc. SPIE}, volume = {9906}, number = {}, pages = {99063E-99063E-14}, year = {2016}, doi = {10.1117/12.2230012}, URL = { http://dx.doi.org/10.1117/12.2230012}, eprint = {} }

@INPROCEEDINGS{Marchiori:2024,
       author = {{Marchiori}, G. and {De Lorenzi}, S. and {Martinez}, J. and others},
        title = "{Rubin Observatory rotating enclosure (dome) progress and status}",
    booktitle = {Ground-based and Airborne Telescopes X},
         year = 2024,
       editor = {{Marshall}, Heather K. and {Spyromilio}, Jason and {Usuda}, Tomonori},
       series = {Society of Photo-Optical Instrumentation Engineers (SPIE) Conference Series},
       volume = {13094},
        month = aug,
          eid = {1309404},
        pages = {1309404},
          doi = {10.1117/12.3018132},
       adsurl = {https://ui.adsabs.harvard.edu/abs/2024SPIE13094E..04M}
}

@misc{oss_doc,
  title = {Observatory System Specifications (OSS)}, 
  howpublished ={Charles F. Claver and the LSST Systems Engineering Integrated Project Team, LSE-30 (rel20.1), Latest Revision Date: May 8, 2024},
}
\bibliographystyle{spiebib} 

\end{document}